\documentclass[journal]{IEEEtran}
\usepackage{cite}
\usepackage[hyphens]{url}
\usepackage{amsmath,amssymb,amsfonts}
\usepackage{cleveref}
\usepackage{algorithm}
\usepackage{algorithmic}
\usepackage{graphicx}
\usepackage{textcomp}
\usepackage{xcolor} 
\usepackage{bm}
\usepackage{subcaption}
\usepackage{multirow} % 多行合并
\usepackage{diagbox} % 提供斜线命令
\usepackage{booktabs}
\usepackage{threeparttable}
\usepackage{authblk}
\usepackage{braket}
\usepackage{xspace}
\iffalse
\newcommand{\REVISE}[1]{\color{red!80!black}{#1}\color{black}\xspace}

\else
\newcommand{\REVISE}[1]{#1\xspace}

\fi

\Crefname{figure}{Fig.}{Figs.}
\Crefname{equation}{Eq.}{Eqs.}
\Crefname{algorithm}{Alg.}{Algs.}

\def\BibTeX{{\rm B\kern-.05em{\sc i\kern-.025em b}\kern-.08em
    T\kern-.1667em\lower.7ex\hbox{E}\kern-.125emX}}

\begin{tiny}

\author{
\IEEEauthorblockN{Bin-Han Lu\IEEEauthorrefmark{1}\IEEEauthorrefmark{2}, Peng Wang\IEEEauthorrefmark{1}\IEEEauthorrefmark{2}, Zhao-Yun Chen\IEEEauthorrefmark{3}, 
Huan-Yu Liu\IEEEauthorrefmark{1}\IEEEauthorrefmark{2}, Tai-Ping Sun\IEEEauthorrefmark{1}\IEEEauthorrefmark{2}, Peng Duan\IEEEauthorrefmark{1}\IEEEauthorrefmark{2},
Yu-Chun Wu\IEEEauthorrefmark{1}\IEEEauthorrefmark{2}\IEEEauthorrefmark{3}, Guo-Ping Guo\IEEEauthorrefmark{1}\IEEEauthorrefmark{2}\IEEEauthorrefmark{3}\\}
\IEEEauthorblockA{
\IEEEauthorrefmark{1} Key Laboratory of Quantum Information Chinese Academy of Sciences, School of Physics, 
University of Science and Technology of China, Hefei, Anhui, 230026, P. R. China\\
\IEEEauthorrefmark{2} CAS Center For Excellence in Quantum Information and Quantum Physics, 
University of Science and Technology of China, Hefei, Anhui, 230026, P. R. China\\
\IEEEauthorrefmark{3} Institute of Artificial Intelligence, Hefei Comprehensive National Science Center, Hefei, Anhui, 230026, P. R. China}
}

\end{tiny}

\title{CAMEL: Physically Inspired Crosstalk-Aware Mapping and gatE scheduLing for Frequency-Tunable Quantum Chips}
\begin{document}
\maketitle
\thispagestyle{plain}
\pagestyle{plain}

%%%%%% -- PAPER CONTENT STARTS-- %%%%%%%%
\begin{abstract}
Crosstalk poses a significant challenge in quantum computing, particularly when quantum gates are executed in parallel, as qubit frequency resonance can lead to residual coupling and reduced gate fidelity. Current solutions struggle to mitigate both crosstalk and decoherence during parallel two-qubit gate operations on frequency-tunable quantum chips. To address this, we propose a crosstalk-aware mapping and gate scheduling (CAMEL) approach, designed to mitigate crosstalk and suppress decoherence by leveraging the tunable coupler’s physical properties and incorporating a pulse compensation technique. CAMEL operates within a two-step compilation framework: first, a qubit mapping strategy that considers both crosstalk and decoherence; and second, a gate timing scheduling method that prioritizes the execution of the largest possible set of crosstalk-free parallel gates, reducing overall circuit execution time. Evaluation results demonstrate CAMEL’s superior ability to mitigate crosstalk compared to crosstalk-agnostic methods, while successfully suppressing decoherence where other approaches fail. Additionally, CAMEL exhibits better performance than dynamic-frequency-aware techniques, particularly in low-complexity hardware environments.
% Crosstalk represents a formidable obstacle in quantum computing.
% When quantum gates are executed parallelly, the resonance of qubit frequencies can lead to residual coupling, compromising the fidelity.
% Existing crosstalk solutions encounter difficulties in mitigating crosstalk and decoherence when dealing with parallel two-qubit gates in frequency-tunable quantum chips.
% Inspired by the physical properties of frequency-tunable quantum chips, we introduce a \textbf{C}rosstalk-\textbf{a}ware \textbf{m}apping and gat\textbf{e} schedu\textbf{l}ing (CAMEL) approach to address these challenges. 
% CAMEL aims to mitigate the crosstalk of parallel two-qubit gates and suppress decoherence.
% Utilizing the features of the tunable coupler, the CAMEL approach integrates a pulse compensation method for crosstalk mitigation.
% Furthermore, we present a compilation framework, including two steps. Firstly, we devise a qubit mapping approach that accounts for both crosstalk and decoherence. 
% Secondly, we introduce a gate timing scheduling approach capable of prioritizing the execution of the largest set of crosstalk-free parallel gates to shorten quantum circuit execution times.
% Evaluation results demonstrate the effectiveness of CAMEL in mitigating crosstalk compared to crosstalk-agnostic methods. 
% Furthermore, in contrast to approaches serializing crosstalk gates, CAMEL successfully suppresses decoherence. 
% Finally, CAMEL exhibits better performance over dynamic-frequency awareness in low-complexity hardware.
\end{abstract}

\section{Introduction}
% Quantum computing has entered the noisy intermediate-scale quantum (NISQ) era, characterized by chips containing dozens to hundreds of qubits \cite{preskill2018quantum}. 
% The superconducting qubit is a promising system for realizing NISQ chips \cite{krantz2019quantum}.
% Superconducting qubits face challenges from two sources of noise. Decoherence, originating from environmental interaction, causes quantum state to lose coherence \cite{bylander2011noise, ithier2005decoherence, yan2016flux}. 
% Crosstalk, resulting from undesired qubit coupling, becomes prominent as qubit frequencies approach resonance, amplifying error during the parallel execution of multiple quantum gates \cite{krinner2020benchmarking}.
% However, NISQ chips are not yet scalable enough for fault-tolerant computing \cite{nielsen2002quantum,neill2018blueprint,maslov2021quantum}. 
% Consequently, effective noise suppression approaches are imperative \cite{almudever2017engineering}.

% It is crucial to ensure that quantum program execution completes before qubit states experience full decoherence.
% Proposed solutions \cite{siraichi2019qubit, ash2019qure, siraichi2018qubit} optimize qubit mapping to reduce execution times. 
% However, the presence of crosstalk implies that parallelism can significantly amplify error. If quantum gates must be serialized to mitigate crosstalk, the time optimization benefits of the mapping approach would be substantially diminished.
Quantum computing has advanced into the noisy intermediate-scale quantum (NISQ) era, characterized by chips containing dozens to hundreds of qubits \cite{preskill2018quantum}. Superconducting qubits are among the leading technologies for NISQ devices \cite{krantz2019quantum}, but despite their potential, they face significant challenges from two primary sources of noise: decoherence and crosstalk. Decoherence, caused by environmental interactions, gradually erodes quantum coherence over time \cite{bylander2011noise, ithier2005decoherence, yan2016flux}. Meanwhile, crosstalk, resulting from unintended qubit coupling, becomes problematic when qubit frequencies approach resonance, leading to increased errors during the parallel execution of multiple quantum gates \cite{krinner2020benchmarking}. These noise-related issues greatly affect the performance and reliability of NISQ systems.

Effective noise suppression strategies are essential \cite{almudever2017engineering} to ensure quantum computations are completed before significant decoherence occurs. Current solutions \cite{siraichi2019qubit, siraichi2018qubit} focus on optimizing qubit mapping to reduce execution time. However, crosstalk complicates this, as increased parallelism can amplify errors. Simply serializing gate execution to mitigate crosstalk would negate the time-saving benefits of optimized mapping. Therefore, approaches that address both decoherence and crosstalk while maintaining parallelism are critical for fully realizing the potential of NISQ quantum processors.

% To mitigate crosstalk, software-level proposals involve serializing the execution of parallel gates affected by severe crosstalk. 
% Murali et al. \cite{murali2020software} proposed a scheduling approach considering both crosstalk and decoherence through a multi-objective function. 
% Hua et al. \cite{Hua2022CQCAC} suggested a gate scheduling approach based on graph coloring, while Xie et al. \cite{xie2021mitigating} proposed an instruction reordering approach based on commutativity. 
% These approaches adopt the strategy of serializing parallel gates affected by crosstalk. However, this extends the execution time of the quantum circuit, increasing the risk of decoherence error.
% Furthermore, researchers are investigating hardware-level methods to mitigate crosstalk. 
% These methods include optimizing qubit architecture \cite{finck2021suppressed, ku2020suppression}, employing microwave control for qubit and coupler \cite{ni2022scalable, wei2021quantum}, 
% and adjusting tunable coupler to a ZZ crosstalk minimum frequency \cite{zhao2021suppression, mundada2019suppression}. These approaches effectively address the crosstalk issue among single-qubit gates.
To mitigate crosstalk, software-level solutions often involve serializing the execution of parallel gates significantly impacted by crosstalk. A scheduling approach has been proposed that accounts for both crosstalk and decoherence by employing a multi-objective function \cite{murali2020software}. Another solution introduces gate scheduling based on graph coloring to reduce crosstalk during parallel gate execution \cite{Hua2022CQCAC}. While these approaches address crosstalk by serializing affected gates, this increases the overall execution time of quantum circuits, thereby raising the risk of decoherence errors.

In addition to software solutions, researchers are exploring hardware-level methods to reduce crosstalk. These efforts include optimizing qubit architecture \cite{finck2021suppressed, ku2020suppression}, using microwave control for qubits and couplers \cite{ni2022scalable, wei2021quantum}, and adjusting tunable couplers to reach a minimum ZZ crosstalk frequency \cite{zhao2021suppression, mundada2019suppression}. These hardware approaches are particularly effective at mitigating crosstalk in single-qubit gates.

% However, in frequency-tunable quantum chips \cite{sung2021realization}, during the execution of two-qubit gates, 
% the qubit frequencies shift from the single-qubit (idle) frequency to a two-qubit (interaction) frequency.
% If the neighboring (spectator) qubit's frequency is near-resonant with that of the gate qubit, unwanted population swap can occur, making previous hardware solutions ineffective. 
% This phenomenon is referred to as frequency crowding.
% To address this, researchers proposed a frequency configuration technique to reduce crosstalk by avoiding near-resonance \cite{klimov2024optimizing}. 
% However, with increasing chip scale and the need for arbitrary parallel gate execution, the exponential rise in parallel two-qubit gate scenarios makes frequency configuration impractical.
% As an alternative, Ding et al. \cite{Ding2020SystematicCM} proposed a dynamic frequency configuration approach. 
% However, this approach lacks calibration of gate pulse parameters. 
% This process, referred to as calibration, involves repeatedly executing gates and fine-tuning their control parameters to minimize errors \cite{wittler2021integrated}.
% Calibration must be completed before circuit execution \cite{klimov2020snake}. 
% Implementing real-time calibration for the dynamic frequency configuration strategy during circuit execution, required by \cite{Ding2020SystematicCM}, is impossible.

In frequency-tunable superconducting qubits~\cite{sung2021realization}, qubit frequencies shift from their single-qubit idle frequency to a two-qubit interaction frequency during the execution of two-qubit gates. If the frequency of a neighboring spectator qubit is near resonance with the gate qubit, unwanted population swap can occur, making previous hardware solutions ineffective. This issue is referred to as frequency crowding. To address this, a frequency configuration technique was proposed to reduce crosstalk by avoiding near-resonance scenarios \cite{klimov2024optimizing}. However, as the scale of chips increases and arbitrary parallel gate execution becomes necessary, the exponential growth of parallel two-qubit gate scenarios renders frequency configuration impractical. As an alternative, Ding et al. \cite{Ding2020SystematicCM} proposed a dynamic frequency configuration approach. However, this method does not include the calibration of gate pulse parameters, which involves repeatedly executing gates and fine-tuning control parameters to minimize errors \cite{wittler2021integrated}. Calibration must be completed prior to circuit execution \cite{klimov2020snake}. Implementing real-time calibration for the dynamic frequency configuration strategy, as required by~\cite{Ding2020SystematicCM}, is not feasible during circuit execution.

Building on the physical properties of frequency-tunable superconducting qubits, we introduce a scalable approach: \textbf{c}rosstalk-\textbf{a}ware \textbf{m}apping and gat\textbf{e} schedu\textbf{l}ing (CAMEL). This approach is designed to mitigate both crosstalk and decoherence in frequency-tunable systems. We have demonstrated that crosstalk errors in two-qubit gates on frequency-tunable quantum chips stem from population swap between gate qubits and spectator qubits. Furthermore, we identified a population swap cutoff frequency in the spectator coupler, indicating that near-resonance effects between two-qubit gates and spectator qubits can still be mitigated by applying compensation pulses to the spectator coupler, even after frequency configuration has been completed.

\REVISE{To further extend crosstalk mitigation beyond local windows, we introduce a crosstalk-aware mapping strategy along with a gate scheduling method based on the maximum independent set problem, optimizing execution time while effectively reducing crosstalk across the chip.}
To address the challenge posed by the growing number of parallel gate execution scenarios, we partition the chip into local windows. Compared to the dynamic frequency configuration approach proposed in \cite{Ding2020SystematicCM}, CAMEL completes calibration prior to quantum program execution, as the number of windows and qubits remains consistent.

% \CZY{Remove this paragraph?} 
% First, we employ a mapping strategy that takes crosstalk into account, aiming to map parallel two-qubit gates to non-crosstalk windows to simultaneously mitigate both decoherence and crosstalk. However, due to the heuristic nature of this method, fully eliminating crosstalk is still not feasible. Second, we introduce a gate scheduling approach based on the maximum independent set problem from graph theory \cite{hagberg2020networkx}. We model crosstalk among gates as a graph and solve the maximum independent set problem to prioritize executing the largest possible set of crosstalk-free gates. This method maximizes the parallel execution of two-qubit gates, thereby reducing quantum program execution times while effectively mitigating crosstalk.

% Our contributions can be summarized as follows:
% \begin{itemize}
% \item We introduced a pulse compensation method to mitigate two-qubit gate population swap crosstalk in frequency-tunable quantum chips, 
% proving its effectiveness through theoretical analysis and simulations.
% \item \REVISE{We propose a qubit mapping and gate scheduling approach that extends the local crosstalk suppression capability of the compensation pulse to the entire chip.}
% \item During the evaluation phase, we used commonly employed benchmarks in the current NISQ era and cross entropy benchmarking circuits \cite{arute2019quantum}, 
% all of which yielded satisfactory results.
% \end{itemize}
Our key contributions are as follows:
\begin{itemize}
\item A pulse compensation method was introduced to mitigate two-qubit gate population swap crosstalk in frequency-tunable superconducting qubits, and its effectiveness was demonstrated through both theoretical analysis and simulations.
\item A crosstalk-aware qubit mapping and gate scheduling approach was developed, expanding the local crosstalk suppression capability of the compensation pulse across the entire chip.
\item The evaluation was performed using widely adopted benchmarks for the current NISQ era, including cross-entropy benchmarking circuits \cite{arute2019quantum}, all of which produced satisfactory results.
\end{itemize}

\section{Preliminaries}
% \subsection{Basic information for quantum computing}
% \subsubsection{Qubit, gate and quantum circuit}
\subsection{Fundamental information for quantum computing}
\REVISE{
A qubit, the fundamental unit of quantum information, is described by a superposition $|\psi\rangle = \alpha|0\rangle + \beta|1\rangle$, where $|\alpha|^2 + |\beta|^2 = 1$. 
A quantum computer with $n$ qubits has a superposition over $2^n$ basis states. 
Quantum computations are executed with quantum gates, which can be decomposed into single or two-qubit gates. 
% Quantum circuits, consisting of input qubits, quantum gates, measurements, and classical registers, serve as the building blocks of quantum programs~\Cref{fig circuit dag}(a). 
Quantum circuits, consisting of input qubits, quantum gates, measurements, and classical registers, serve as the building blocks of quantum programs. 
% The circuit can also be modeled as a directed acyclic graph (DAG) $\mathcal{D}(\bm{q},\bm{g},\bm{e})$ \Cref{fig circuit dag}(b)~\cite{Clarktdag}.} 
The circuit can also be modeled as a directed acyclic graph (DAG) $\mathcal{D}(\bm{q},\bm{g},\bm{e})$~\cite{Clarktdag}.} 
\REVISE{The coupling structure of a quantum chip is represented as a graph $\mathcal{G}(\bm{Q}, \bm{E})$, where $\bm{Q}$ denotes the physical qubits and $\bm{E}$ represents the couplers. Two-qubit gates (such as CZ gate) between $q_i\in \bm{Q}$ and $q_j \in \bm{Q}$ can be executed only when $(q_i, q_j)\in \bm{E}$.}

Quantum systems are highly sensitive to environmental noise, which can induce decoherence~\cite{ithier2005decoherence}.
The probability of qubit decoherence increases exponentially over time, governed by $P(t) = 1 - \exp\left(-\frac{t}{T_i}\right)$,
where $T_i$ represents the decoherence time. 
There are two types of decoherence times: the relaxation time $T_1$, 
which represents the transition rate from the excited state to the ground state, and the dephasing time $T_2$, which represents the rate of phase information loss. 
To prevent errors, a quantum circuit must complete its execution within the decoherence time.

\subsection{Superconducting quantum computing hardware}
% \subsubsection{Decoherence}

\subsubsection{Tunable qubit}\label{subsubsect tunable qubit}

\begin{figure*}[htbp]
\begin{minipage}[t]{0.24\linewidth}
\centering
\includegraphics[width=\textwidth]{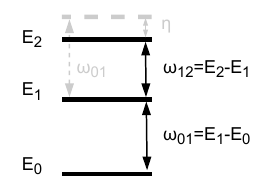}\\
(a)
\end{minipage}
\begin{minipage}[t]{0.24\linewidth}
\centering
\includegraphics[width=\textwidth]{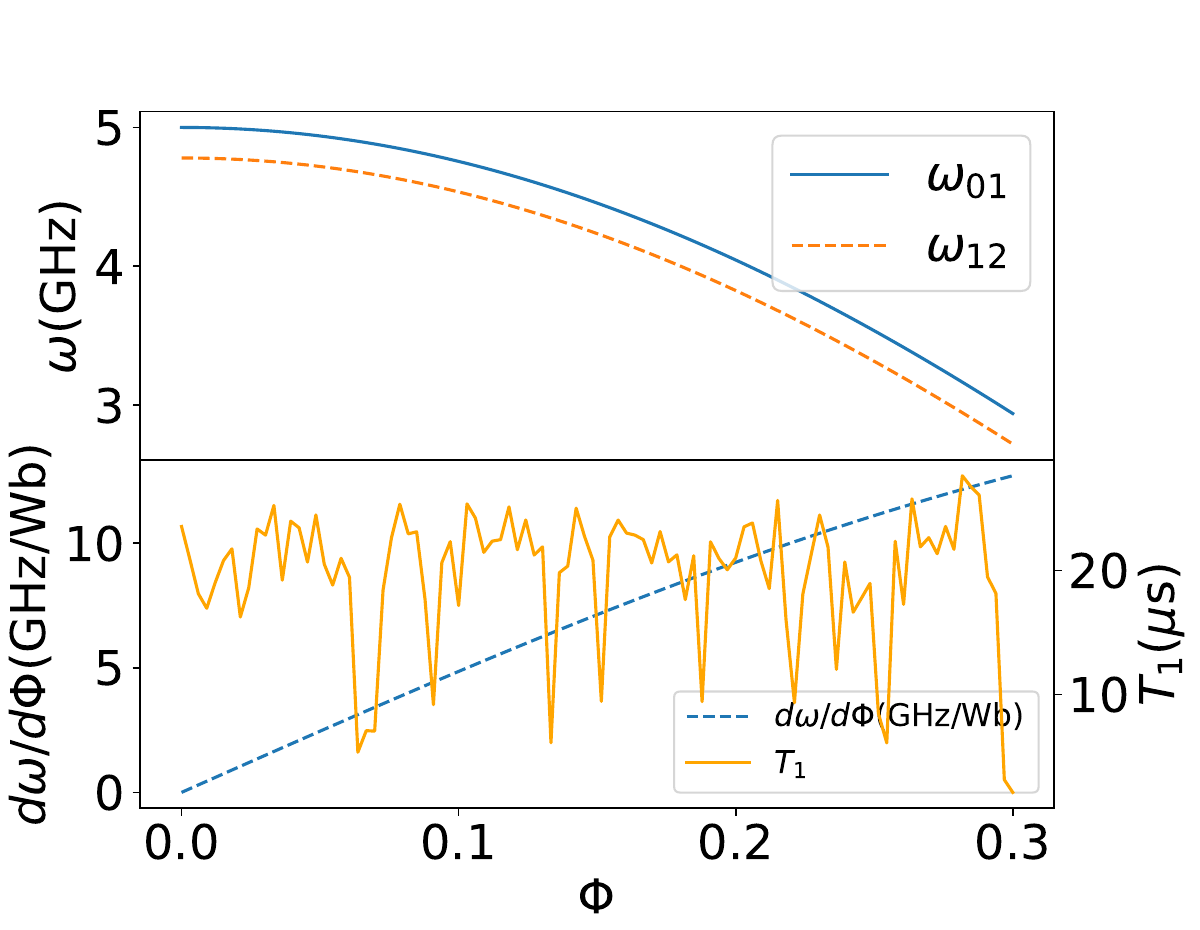}\\
(b)
\end{minipage}
\begin{minipage}[t]{0.24\linewidth}
\centering
\includegraphics[width=\textwidth]{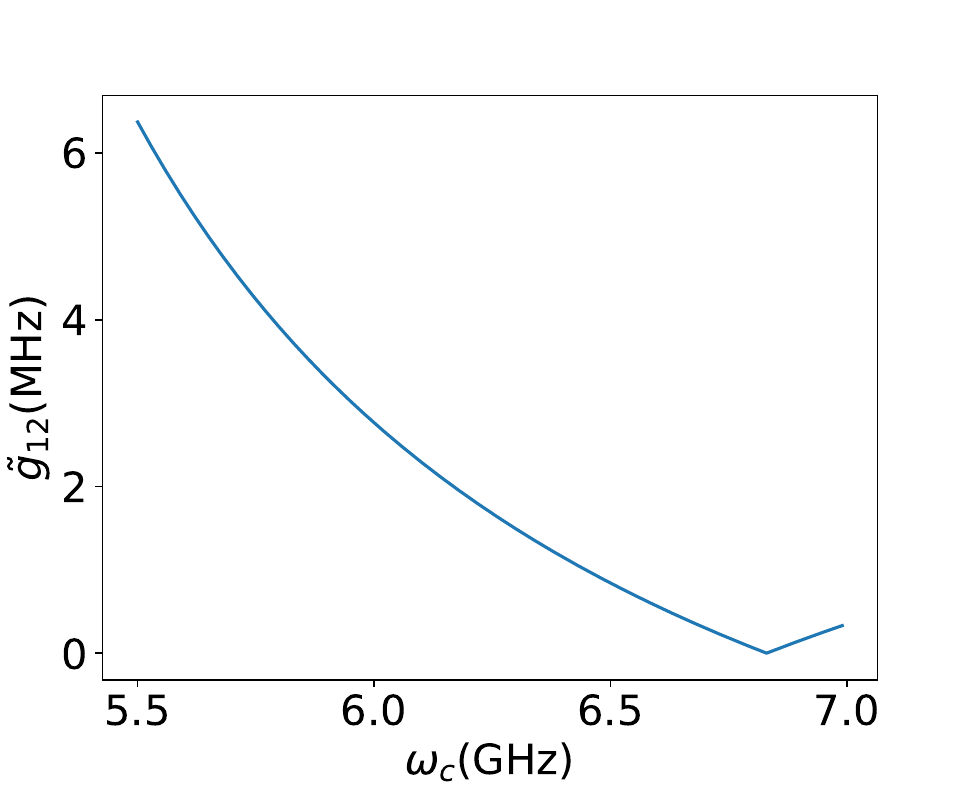}\\
(c)
\end{minipage}
\begin{minipage}[t]{0.24\linewidth}
\centering
\includegraphics[width=\textwidth]{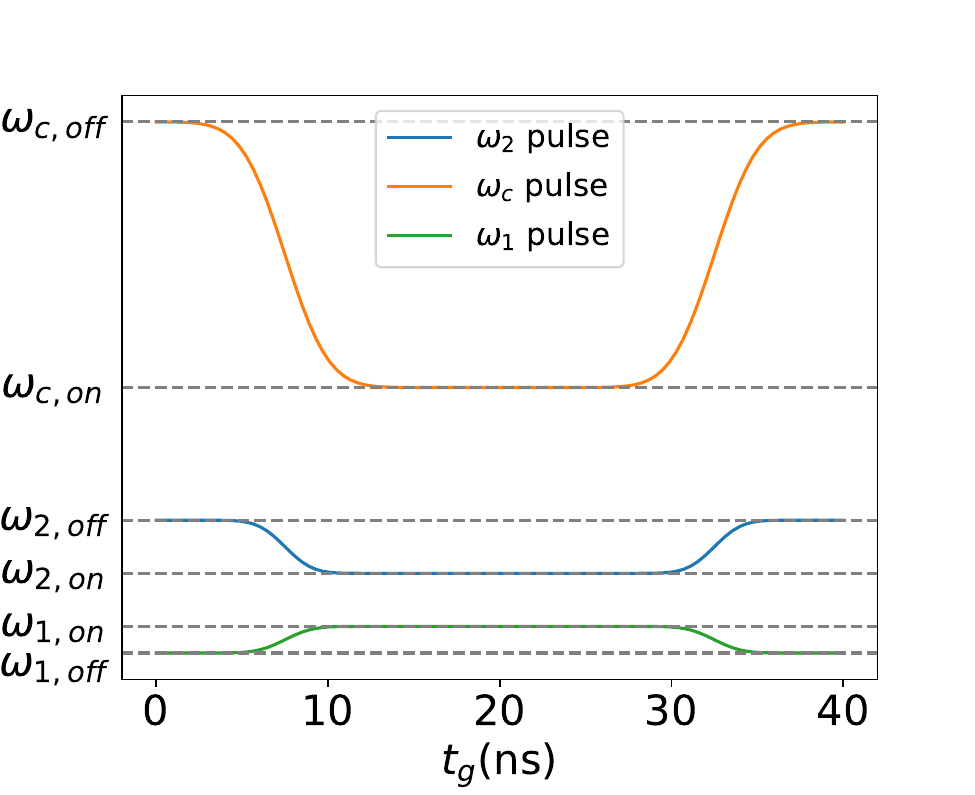}\\
(d)
\end{minipage}
\caption{(a) The energy level diagram for superconducting qubits.
(b) The frequency spectrum for tunable qubits.
When the frequency is close to the maximum point, $d\omega/d\Phi$ is small. $T_1$ is reduced at TLS defect points.
(c) The relationship between $\tilde{g}_{12}$ and the coupler frequency. A specific coupler frequency exists at which $\tilde{g}_{12}$ reaches zero. As the coupler frequency decreases, the magnitude of $|\tilde{g}_{12}|$ increases.
(d) The dynamic gate pulse applied to qubits $Q_1$ and $Q_2$, as well as the coupler $c_{12}$ between them. The qubit frequencies $\omega_i$ shift from $\omega_{i,\text{off}}$ to interaction frequency $\omega_{i,\text{on}}$, 
ensuring the resonant condition $\omega_1 + \eta_1=\omega_2$ is met. Additionally, the coupler frequency decreases from $\omega_{c,\text{off}}$ to $\omega_{c,\text{on}}$ to enlarge $\tilde{g}_{12}$.
}
\label{fig tunable}
\end{figure*}

% \REVISE{The qubit frequency represents the energy transition between the ground state $|0\rangle$ and the excited state $|1\rangle$, with $\omega = \omega_{01} = E_1 - E_0$, 
% as shown in \Cref{fig tunable qubit}(a). The anharmonicity parameter $\eta$ ensures that $\omega_{12} = \omega_{01} + \eta$, 
% detuning from $\omega_{01}$ to prevent transitions to the second excited state $|2\rangle$. In frequency-tunable superconducting qubits (or tunable qubits), the frequency is adjusted by an external magnetic flux $\Phi$, with a frequency spectrum illustrated in \Cref{fig tunable qubit}(b). 
% The $T_1$ is reduced at two-level system (TLS) defect points \cite{muller2015interacting}, while $T_2$ is proportional to $d\omega/d\Phi$ \cite{bylander2011noise}.
% To achieve a long $T_i$, the frequency must be in a region where $d\omega/d\Phi$ is small and away from TLS defect points, imposing constraints on the frequency range.}
\REVISE{The qubit frequency corresponds to the energy transition between the ground state $|0\rangle$ and the excited state $|1\rangle$, represented by $\omega = \omega_{01} = E_1 - E_0$, as shown in \Cref{fig tunable}(a). The anharmonicity parameter $\eta$ ensures that $\omega_{12} = \omega_{01} + \eta$, detuning the second excited state $|2\rangle$ from $\omega_{01}$ to prevent unwanted transitions. In frequency-tunable superconducting qubits (or tunable qubits), the frequency is controlled by an external magnetic flux $\Phi$, with the frequency spectrum shown in \Cref{fig tunable}(b). The coherence time $T_1$ decreases near two-level system (TLS) defect points \cite{muller2015interacting}, while $T_2$ is proportional to $d\omega/d\Phi$ \cite{bylander2011noise}. To achieve a long $T_i$, the qubit frequency must be positioned where $d\omega/d\Phi$ is minimal and away from TLS defect points, constraining the available frequency range.}

\subsubsection{Tunable coupler \cite{bravyi2011schrieffer}}
\REVISE{The tunable coupler is a promising technological approach for superconducting chips, relying on a specialized structure to control the coupling strength between qubits. This technology has seen widespread adoption~\cite{2011Implementing, 2023Generation} and has been employed in important experiments, including those demonstrating quantum supremacy~\cite{arute2019quantum} and quantum error correction codes~\cite{google2023suppressing, acharya2024quantum}. These applications underscore the tunable coupler’s significant potential for scaling up quantum computers.}

In \Cref{fig tunable}(c), the effective coupling strength $\tilde{g}_{12}$ reaches a minimum, signifying the elimination of crosstalk.
In \Cref{fig tunable}(d), the frequencies of $Q_1$ and $Q_2$ satisfy $\omega_1+\eta_1=\omega_2$ \cite{krantz2019quantum}, causing the states $|11\rangle$ and $|02\rangle$ to become resonant.
Simultaneously, the frequency of $c_{12}$ must be tuned to a lower point with larger $\tilde{g}_{12}$, as depicted in \Cref{fig tunable}(c), to ensure fast gate execution.
Subsequently, after $t_g=\pi/\tilde{g}_{12}$, the state $|11\rangle$ undergoes a phase shift: $|11\rangle\xrightarrow{e^{i\pi/2}} i|02\rangle\xrightarrow{e^{i\pi}}-|11\rangle$, thus realizing the CZ gate.

% \begin{figure}[htbp]
% \centering
% \begin{minipage}[t]{0.49\linewidth}
% \centering
% \includegraphics[width=\textwidth]{eff_g.pdf}\\
% (a)
% \end{minipage}
% \begin{minipage}[t]{0.49\linewidth}
% \centering
% \includegraphics[width=\textwidth]{gate_pulse.pdf}\\
% (b)
% \end{minipage}
% \caption{
% (a) The relationship between $\tilde{g}_{12}$ and the coupler frequency. A specific coupler frequency exists at which $\tilde{g}_{12}$ reaches zero. As the coupler frequency decreases, the magnitude of $|\tilde{g}_{12}|$ increases.
% (b) The dynamic gate pulse applied to qubits $Q_1$ and $Q_2$, as well as the coupler $c_{12}$ between them. The qubit frequencies $\omega_i$ shift from $\omega_{i,\text{off}}$ to interaction frequency $\omega_{i,\text{on}}$, 
% ensuring the resonant condition $\omega_1 + \eta_1=\omega_2$ is met. Additionally, the coupler frequency decreases from $\omega_{c,\text{off}}$ to $\omega_{c,\text{on}}$ to enlarge $\tilde{g}_{12}$.}
% \label{fig eff g gate pulse}
% \end{figure}

\subsection{Gate pulse calibration \cite{wittler2021integrated}}
Quantum gates are implemented by applying time-dependent pulses to qubits, represented by the function $f(t;\bm{\alpha})$, 
where $\bm{\alpha}$ denotes the control parameters. Before executing a quantum circuit, the gates undergo calibration to determine the optimal parameters $\bm{\alpha}^*$ that maximize the fidelity. 
This iterative process requires multiple executions and measurements, making it impractical to perform during circuit execution.

\subsection{Crosstalk error}
\subsubsection{Single-qubit gate crosstalk}\label{subsubsect single q gate xtalk}
% \REVISE{
% When the qubit frequencies are close to those of neighboring qubits, crosstalk can occur, 
% leading to transitions such as $\ket{01} \leftrightarrow \ket{10}$ and $\ket{11} \leftrightarrow \ket{02}$~\cite{zhao2022quantum}.
% Additionally, the microwave signal for the single-qubit gate on qubit $Q_i$ may affect non-target qubit $Q_j$ \cite{zhao2022spurious}. 
% To minimize crosstalk, qubits should be in a far-detuned regime. These mechanisms impose constraints on the allowable frequency range.}
\REVISE{ When qubit frequencies are close to those of neighboring qubits, crosstalk can occur, resulting in unwanted transitions such as $\ket{01} \leftrightarrow \ket{10}$ and $\ket{11} \leftrightarrow \ket{02}$~\cite{zhao2022quantum}. Additionally, the microwave signal applied to a single-qubit gate on qubit $Q_i$ can affect non-target qubit $Q_j$ \cite{zhao2022spurious}. To minimize crosstalk, qubits should be kept in a far-detuned regime, imposing constraints on the allowable frequency range.}

\begin{figure}[htbp]
\centering
\begin{minipage}[t]{0.45\linewidth}
\centering
\includegraphics[width=\textwidth]{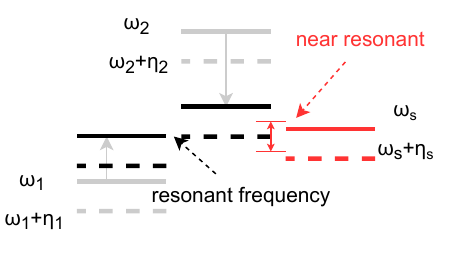}\\
(a)
\end{minipage}
\begin{minipage}[t]{0.3\linewidth}
\centering
\includegraphics[width=\textwidth]{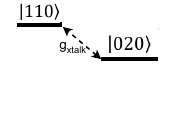}\\
(b)
\end{minipage}
\caption{
(a) The energy level model for the CZ gate qubits $Q_1$ and $Q_2$, alongside the spectator qubit $Q_s$. $\omega_1$ and $\omega_2 + \eta_2$ denote interaction frequencies.
The frequency $\omega_s$ of the spectator qubit is close to the resonant frequency.
\REVISE{
(b) The crosstalk model considers the qubits in the order $|Q_sQ_2Q_1\rangle$, where $Q_1$ and $Q_2$ are interacting in state $\ket{020}$. 
If $E_{110}$ resonates with $E_{020}$, population swap from $\ket{020}$ to $\ket{110}$ will occur.}
}
\label{fig spectator}
\end{figure}

% When the frequencies of neighboring qubits are in the far-detuned regime, there is no population swap between them, and the coupling Hamiltonian is given by $H/\hbar=E_{\widetilde{n}}\ket{\widetilde{n}}$. 
% Here, $E_{\widetilde{n}}$ represents dressed state energy level, 
% and the energy shift $\xi=E_{\widetilde{11}}-E_{\widetilde{01}}-E_{\widetilde{10}}+E_{\widetilde{00}}$ is known as ZZ coupling \cite{Xie2022SuppressingZC}. 
% Adjusting the coupler frequency to the ZZ coupling is minimum can effectively suppress this type of crosstalk.
When neighboring qubit frequencies are in the far-detuned regime, there is no population swap between them, and the coupling Hamiltonian is given by $H/\hbar=E_{\widetilde{n}}\ket{\widetilde{n}}$. Here, $E_{\widetilde{n}}$ represents the energy levels of the dressed states, and the energy shift $\xi=E_{\widetilde{11}}-E_{\widetilde{01}}-E_{\widetilde{10}}+E_{\widetilde{00}}$ is referred to as ZZ coupling \cite{Xie2022SuppressingZC}. Adjusting the coupler frequency to minimize the ZZ coupling effectively suppresses this type of crosstalk.

\subsubsection{Two-qubit gate crosstalk}\label{subsubsect two q gate xtalk}
% Let's consider a model in \Cref{fig spectator}(a). $Q_1$ and $Q_2$ are CZ gate qubits, and their frequencies are tuned to the resonant frequency. $Q_s$ is a spectator qubit that is connected with $Q_2$. The quantum state can be represented as follows:
% \begin{equation}
% \begin{aligned}
% |\Psi(t)\rangle&=\cos\tilde{g}_{12}t|011\rangle-i\sin\tilde{g}_{12}t|020\rangle\\
% &=\frac{1}{\sqrt{2}}\left[\cos\tilde{g}_{12}t(|+\rangle+|-\rangle)-i\sin\tilde{g}_{12}t(|+\rangle-|-\rangle)\right]\\
% &=\frac{1}{\sqrt{2}}(e^{-i\tilde{g}_{12}t}|+\rangle+e^{i\tilde{g}_{12}t}|-\rangle),\\
% \end{aligned}
% \end{equation}
% where $E_{011}=\omega_1+\omega_2,E_{110}=\omega_s+\omega_2$.
% If the execution time of a CZ gate is $t_g$, and the effective coupling between qubits $Q_1$ and $Q_2$ is $\tilde{g}_{12}=\pi/t_g$, 
% then the energy levels corresponding to $|\pm\rangle$ are $E_{\pm}=E_{011}\pm\tilde{g}_{12}$. 
% In principle, it's necessary to tune the $E_{110}$ far from $E_{\pm}$ to avoid population swap.
% \REVISE{This physical mechanism also adds a constraint on the frequencies for both single-qubit and two-qubit gates.}

\REVISE{
There are various scenarios of frequency crowding between the two-qubit gates and the spectator qubit. 
We take one of these scenarios as an example to explain in detail the physical mechanism of the population swap from the gate state to the excited state of the spectator qubit.
Consider the model in \Cref{fig spectator}(a), where qubits $Q_1$ and $Q_2$  serve as the CZ gate qubits, with their frequencies tuned to resonance, and  $Q_s$ acts as a spectator qubit connected to $Q_2$. The energy levels satisfy $E_{020}=E_{011}$, which correspond to the CZ gate level. Additionally, $E_{011} = \omega_1 + \omega_2$ and  $E_{110} = \omega_s + \omega_2$. To prevent crosstalk, it is crucial to ensure that $E_{110}$ is sufficiently far from $E_{020}$ during the execution of the CZ gate.
Otherwise, a population swap from $\ket{020}$ to $\ket{110}$ will occur \Cref{fig spectator}(b).
}

\subsubsection{Frequency crowding}
% \REVISE{The constraints in \Cref{subsubsect tunable qubit,subsubsect single q gate xtalk,subsubsect two q gate xtalk} compress the feasible frequency range. 
% Deviations from the sweet spot (around 500 MHz) significantly shorten the decoherence time, so the frequency should remain within 
% $(\omega_\text{max}-500 \text{MHz}, \omega_\text{max})$ \cite{klimov2020optimizing}. 
% On a square periodic chip, each qubit has 4 nearest neighbors and 8 next-nearest neighbors, 
% while each two-qubit gate is affected by 6 spectator qubits and 10 potentially crosstalking parallel gates. 
% Each near-resonance region exceeds 20 MHz, further reducing the effective frequency range. 
% If two parallel gates experience crosstalk due to qubit frequency crowding, they must be executed sequentially. 
% However, this adjustment extends the quantum circuit runtime, increasing the risk of decoherence. We will address how to mitigate this issue, known as frequency crowding, in the next section.}

\REVISE{The constraints discussed in the previous sections reduce the available frequency range for qubits. Moreover, straying too far from the optimal ``sweet point'' (around 500 MHz) significantly shortens the decoherence time, meaning the qubit frequency must stay within the range $(\omega_\text{max} - 500 \text{MHz}, \omega_\text{max})$ to maintain coherence \cite{klimov2020optimizing}. On a square periodic chip, each qubit has four nearest neighbors and eight next-nearest neighbors, while each two-qubit gate is affected by six spectator qubits and up to 10 potentially crosstalking parallel gates. Each near-resonance region spans over 20 MHz, further limiting the effective frequency range, which is known as ``frequency crowding''. If two parallel gates experience crosstalk due to qubit frequency crowding, they must be executed sequentially. This adjustment, however, increases circuit runtime and raises the risk of decoherence.}

\section{Using Compensation Pulse to Mitigate Crosstalk}\label{sect compensation pulse}
\subsection{Compensation pulse}

\begin{figure*}[htbp]
\centering
\begin{minipage}[t]{0.3\linewidth}
\centering
\includegraphics[width=\textwidth]{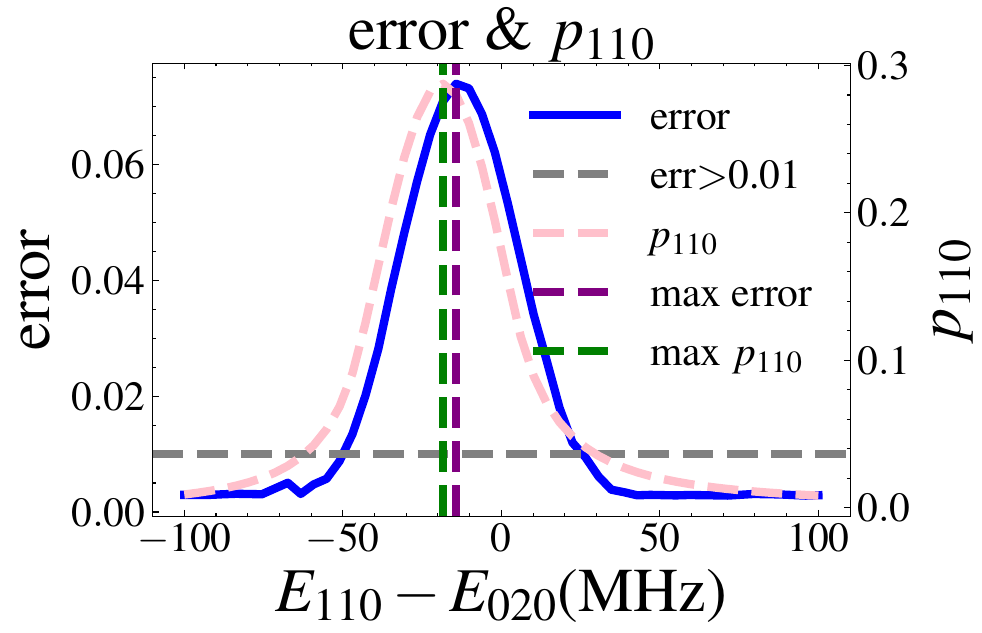}\\
(a)
\end{minipage}
\begin{minipage}[t]{0.25\linewidth}
\centering
\includegraphics[width=\textwidth]{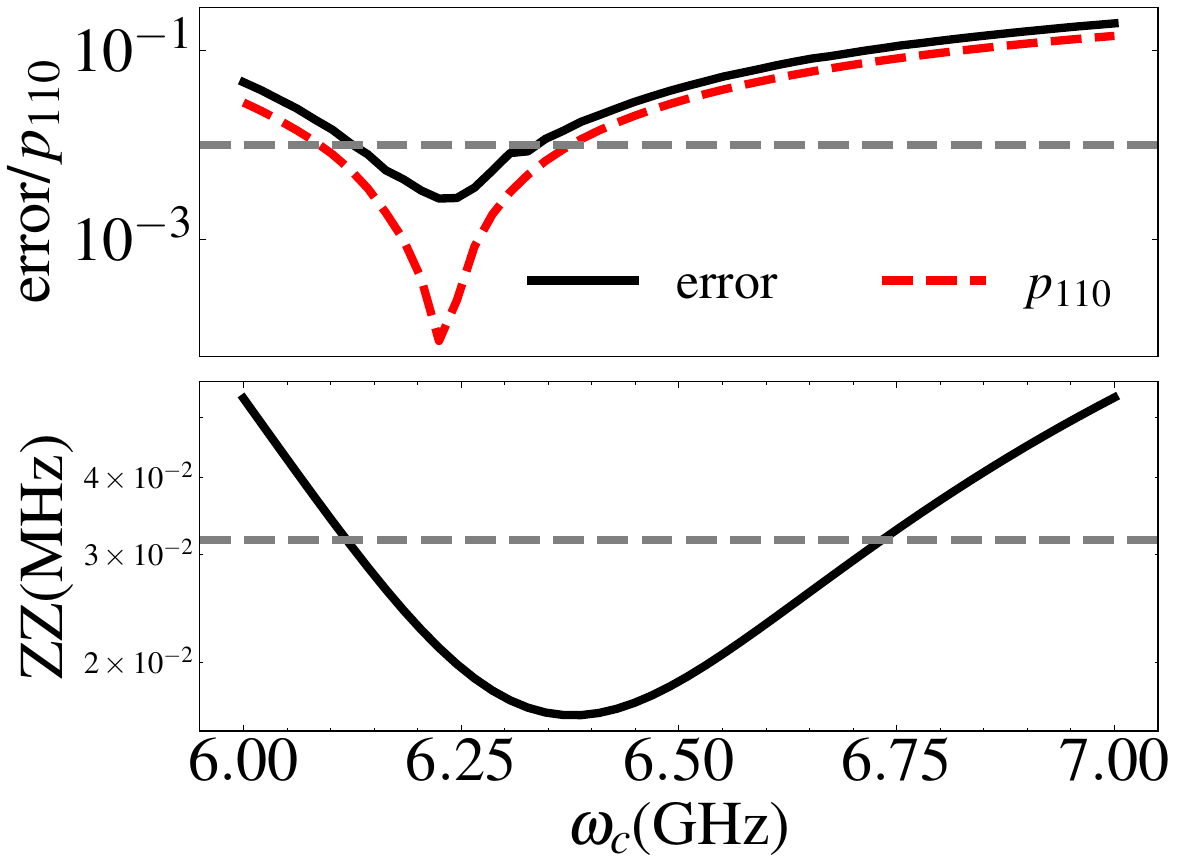}\\
(b)
\end{minipage}
\begin{minipage}[t]{0.28\linewidth}
\centering
\includegraphics[width=\textwidth]{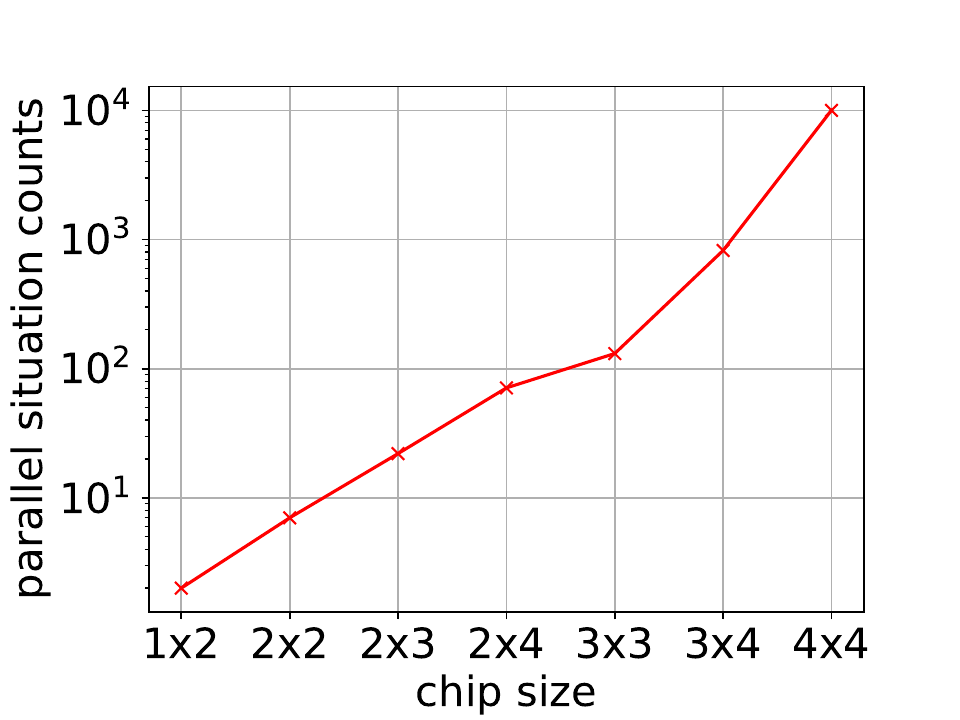}\\
(c)
\end{minipage}
\caption{
\REVISE{(a) The population swap from $\ket{020}$ to $\ket{110}$ and gate errors when the spectator qubit is at different frequencies. The horizontal axis represents the energy difference between 
$|110\rangle$ and $\ket{020}$. As the frequency of the spectator qubit varies, both population swap and gate errors clearly increase simultaneously. In the 100 MHz range, gate errors exceed 0.01, indicated by the grey line.
(b) The relationship between CZ gate error, population swap $p_{110}$ and ZZ coupling on coupler frequency, with the qubit order as $|Q_sQ_2Q_1\rangle$.
(c) The diagram illustrates how the number of parallel situations for two-qubit gates changes with the scale of the $M \times N$ chip.
}
}\label{fig leakage}
\end{figure*}
% \begin{figure}[htbp]
% \centerline{
% \includegraphics[width=0.4\textwidth]{errandleakage.pdf}
% }
% \caption{
% \REVISE{
% This figure shows the population swap from $\ket{020}$ to $\ket{110}$ and gate errors when the spectator qubit is at different frequencies. The horizontal axis represents the energy difference between 
% $|110\rangle$ and $\ket{020}$. As the frequency of the spectator qubit varies, both population swap and gate errors clearly increase simultaneously. In the 100 MHz range, gate errors exceed 0.01, indicated by the grey line.}
% }
% \label{fig errleak}
% \end{figure}  
\REVISE{
In \Cref{subsubsect two q gate xtalk}, we hypothesize that the crosstalk is caused by the population swap from the gate state $\ket{020}$ to the spectator qubit's excited state $|110\rangle$. 
To investigate this, we conducted simulations by varying the frequency of the spectator qubit. 
In \Cref{fig leakage}(a), we observe that population swap and error increase simultaneously when $E_{110}$ approaches resonance with $E_{020}$.
If this entire range above the grey line is treated as a frequency exclusion zone, 
the frequency constraints on qubits will become too restrictive, resulting in significant frequency crowding. 
Next, we will demonstrate that this population swap can be suppressed by adjusting the coupler frequency.

% \begin{figure}[htbp]
% \centering
% \begin{minipage}[t]{0.53\linewidth}
% \centering
% \includegraphics[width=\textwidth]{errandleakage.pdf}\\
% (a)
% \end{minipage}
% \begin{minipage}[t]{0.44\linewidth}
% \centering
% \includegraphics[width=\textwidth]{leakage.pdf}\\
% (b)
% \end{minipage}
% \caption{
% \REVISE{(a) This figure shows the population swap from $\ket{020}$ to $\ket{110}$ and gate errors when the spectator qubit is at different frequencies. The horizontal axis represents the energy difference between 
% $|110\rangle$ and $\ket{020}$. As the frequency of the spectator qubit varies, both population swap and gate errors clearly increase simultaneously. In the 100 MHz range, gate errors exceed 0.01, indicated by the grey line.
% (b) The relationship between CZ gate error, population swap $p_{110}$ and ZZ coupling on coupler frequency, with the qubit order as $|Q_sQ_2Q_1\rangle$.}
% }\label{fig leakage}
% \end{figure}

As a result of the unwanted coupling between of $Q_2$ and $Q_s$, the system Hamiltonian is expressed as follows:

\begin{equation}
H/\hbar=
\begin{pmatrix}
E_{020}/\hbar&g_\text{xtalk}\\
g_\text{xtalk}&E_{110}/\hbar
\end{pmatrix}.\label{eq h three}
\end{equation}
% where
% \begin{equation}\label{eq gp}
% \begin{aligned}
% g_+=&g_{2s}+\frac{g_{2c}g_{sc}}{2}\left(\frac{1}{\omega_1+\sqrt{2}\tilde{g}_{12}-\omega_c}+\frac{1}{\omega_s-\omega_c}-\right.\\
% &\left.\frac{1}{\omega_c+\omega_1}-\frac{1}{\omega_c+\omega_s-\sqrt{2}\tilde{g}_{12}}\right),\\
% g_-=&-g_{2s}-\frac{g_{2c}g_{sc}}{2}\left(\frac{1}{\omega_1-\sqrt{2}\tilde{g}_{12}-\omega_c}+\frac{1}{\omega_s-\omega_c}-\right.\\
% &\left.\frac{1}{\omega_c+\omega_1}-\frac{1}{\omega_c+\omega_s+\sqrt{2}\tilde{g}_{12}}\right).
% \end{aligned}
% \end{equation}
The first energy level corresponds to $\ket{020}$, and the second energy level represents $\ket{110}$. 
The term $g_\text{xtalk}$ denotes the coupling strength between $\ket{020}$ and $\ket{110}$, which is a function of $\omega_{1}$, $\omega_{2}$, $\omega_{s}$, and $\omega_{c}$ \cite{chu2021coupler}.
During the execution of single-qubit gates, the detuning between $Q_s$ and $Q_2$, as well as between $Q_2$ and $Q_1$, is large, and the $\omega_c$ is set at the ZZ minimum, resulting in a negligible $g_\text{xtalk}$. However, when executing two-qubit gates on $Q_1$ and $Q_2$, the qubits are tuned to a resonance frequency, and the coupler moves away from the ZZ minimum.
This change reduces the detuning between the qubits and results in a non-zero $g_\text{xtalk}$. Zajac et al.~\cite{zajac2021spectator} proposed a method using compensation pulses to mitigate stray coupling in fixed-frequency qubits.
Through simulations, we found that in a frequency-tunable system, we can similarly adjust the $\omega_c$ with compensation pulses to suppress the population swap from the gate state $\ket{020}$ to the state $\ket{110}$.}
% \begin{figure}[htbp]
% \centerline{
% \includegraphics[width=0.45\textwidth]{leakage.pdf}
% }
% \caption{
% The relationship between (a) CZ gate error, population swap and (b) ZZ coupling on coupler frequency,
% with the qubit order as $|Q_sQ_2Q_1\rangle$.}
% \label{fig population swap}
% \end{figure}

% \begin{figure}[htbp]
% \centering
% \begin{minipage}[t]{0.4\linewidth}
% \centering
% \includegraphics[width=\textwidth]{compensation.pdf}\\
% (a)
% \end{minipage}
% \begin{minipage}[t]{0.55\linewidth}
% \centering
% \includegraphics[width=\textwidth]{parallel_compensate.pdf}\\
% (b)
% \end{minipage}
% \caption{
% (a) During a gate between $Q_1$ and $Q_2$, a pulse is applied to the coupler spectator coupler $c_{2s}$. The frequency of $c_{2s}$ is adjusted from the minimum ZZ frequency to a new frequency, cutting off crosstalk from the spectator qubit.
% (b) Similarly, pulses are applied to $c_{14}$ and $c_{23}$ during parallel gates.}
% \label{fig compensate}
% \end{figure}

\REVISE{
In the simulation, we set $E_{110}=E_{020}$, shifting the $\omega_c$ to calculate the population swap and error, comparing them with the ZZ coupling strength at the idle frequency.
In \Cref{fig leakage}(b), the frequency corresponding to the minimum error is not the same as the ZZ minimum frequency. Additionally, the minimum population swap between $\ket{020}$ and $\ket{110}$ coincides with the minimum error. 
Therefore, it is possible to readjust the spectator coupler to ensure that $g_\text{xtalk}$ is small, thus protecting the fidelity of the two-qubit gate from the crosstalk of spectator qubit $Q_s$.}

Hence, we propose a pulse compensation approach for the population swap of parallel quantum gates. 
If a frequency crowding occurs between the two-qubit gate and spectator qubits, we dynamically adjust the spectator coupler's frequency. 
This adjustment shifts from the ZZ coupling minimum to the population swap minimum.
Similar to gate calibration, this compensation pulse calibration must be done before circuit execution.
\REVISE{
By using compensation pulses, even when the gate qubits are resonant with the frequency of spectator qubits, population swap can be avoided. 
This approach partially mitigates frequency crowding, reduces the constraints on the frequency range, 
and lessens the difficulty of subsequent quantum circuit mapping and scheduling.}

% \begin{figure}[htbp]
% \centerline{
% \includegraphics[width=0.3\textwidth]{chipsit.pdf}
% }
% \caption{
% \REVISE{The diagram illustrates how the number of parallel situations for two-qubit gates changes with the scale of the $M \times N$ chip.}
% }
% \label{fig chipsit}
% \end{figure}

\begin{figure*}[htbp]
\centerline{
\includegraphics[width=0.8\textwidth]{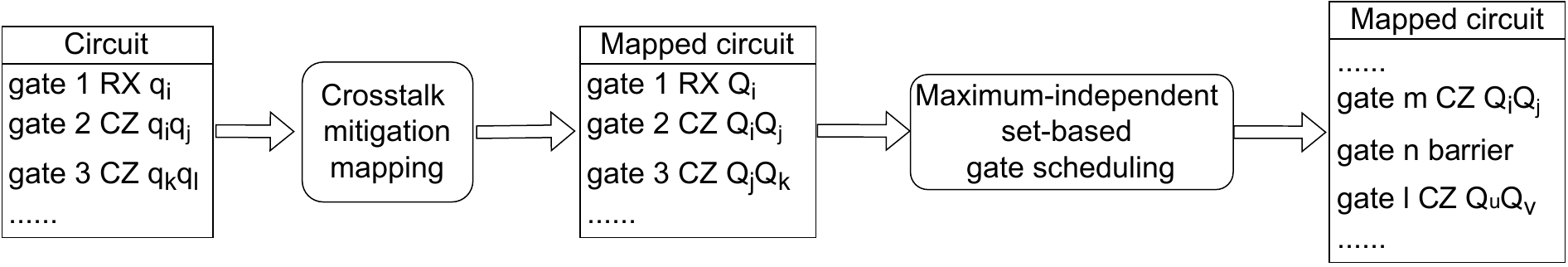}
}
\caption{
The algorithm consists of two main steps. Firstly, given a quantum circuit, it undergoes crosstalk-aware mapping to produce a mapped circuit, where logical qubits $q$ are mapped to physical qubits $Q$. Secondly, gate scheduling based on the maximum independent set is applied, with barriers inserted at appropriate positions to mitigate crosstalk.
}
\label{fig alg overview}
\end{figure*}

\REVISE{
\subsection{Limitation of compensation pulse approach}\label{subsect limit of cp}
Although compensation pulses can mitigate crosstalk errors on a quantum chip, applying them across the entire chip poses a computationally hard problem. In any given quantum circuit, a large number of parallel configurations of two-qubit gates can occur within a single circuit layer. 
Identifying all possible parallel configurations of two-dimensional two-qubit gates on the chip is equivalent to solving an independent set problem, and the number of independent sets in a graph generally increases exponentially with the size of the graph \cite{samotij2015counting,jou2000number}. For a square periodic planar chip structure of size $M \times N$, we calculate the growth in the number of independent sets as $M$ and $N$ increase, as shown in \Cref{fig leakage}(c). Consequently, the compensation pulse cannot achieve global crosstalk suppression across the chip.

The limitation of the compensation pulse technique forces us to focus on mitigating crosstalk in a smaller region of the chip, referred to as a windowed compensation pulse. Developing an effective way to integrate this windowed compensation pulse into a compilation scheme is crucial for reducing quantum circuit errors on a broader scale.
}

 % For a square periodic planar chip structure of size $M \times N$, we calculate the growth in the number of independent sets as $M$ and $N$ increase, as shown in \Cref{fig leakage}(c). Our findings confirm that the number of independent sets—and thus, the number of parallel configurations—grows exponentially. Consequently, the compensation pulse cannot achieve global crosstalk suppression across the chip.

\section{Qubit Mapping \& Gate Scheduling}
% \subsection{Algorithm Overview}
% \REVISE{
% Decoherence necessitates a shorter execution time for a quantum circuit, while frequency crowding requires that parallel quantum gates be executed sequentially, leading to an increase in the overall execution time of the quantum circuit. Additionally, the compensation pulse can only mitigate crosstalk locally. Due to these limitations, an optimal qubit mapping and gate scheduling approach that extends the local crosstalk mitigation capabilities of the compensation pulse to the entire chip is essential.}
% In this section, we will systematically introduce our compilation approach CAMEL, outlining each step to illustrate how our design mitigates crosstalk and decoherence. To provide an overview of our algorithm, we illustrate the algorithmic flow of CAMEL in Fig.~\ref{fig alg overview}.
\REVISE{
Decoherence requires minimizing the execution time of a quantum circuit, while frequency crowding demands that parallel quantum gates be executed sequentially, which in turn increases the overall execution time. Additionally, the compensation pulse can only mitigate crosstalk locally. Given these constraints, it is crucial to develop an optimal qubit mapping and gate scheduling strategy that extends the local crosstalk mitigation capabilities of the compensation pulse across the entire chip.
}
In this section, we systematically introduce our CAMEL compilation approach, detailing each step to demonstrate how our design mitigates both crosstalk and decoherence. To provide an overview, the algorithmic flow of CAMEL is illustrated in Fig.~\ref{fig alg overview}.

\subsection{Basic elements}
\subsubsection{Distance matrix}
Given a coupling graph $\mathcal{G}(\bm{Q},\bm{E})$, we define a distance matrix $D(\cdot,\cdot)$ where each element represents the shortest path between qubit-pairs.
\subsubsection{Top layer}
The top layer, denoted as $F$, consists of pending gates that do not have any unexecuted predecessors within the DAG.
For instance, $g(q_i,q_j)$ is appropriate to be placed in the set $F$ once all preceding gates on $q_i$ and $q_j$ have been executed.
\subsubsection{Gate duration}
Given that the execution time of gate $g$ is $t_g$, if it starts executing at time $g.t$, it finishes execution at $g.t+t_g$.
\subsubsection{Swap gate}
Suppose a mapping at time $t_1$ is denoted by $\pi_1$. If we insert a swap gate $s$, we will obtain a new mapping $\pi_2$
$\pi_1(q_1)=\pi_2(q_2)\cap\pi_1(q_2)=\pi_2(q_1)\cap\pi_1(q)=\pi_2(q), \forall q\neq q_1,q_2$ at $t_2\leftarrow t_1+t_s$.
We define all possible swap gate as a set $\bm{S}$.

\subsection{Constraint}
%Here, we introduce the constraints based on the physical background.
%\subsubsection{Coupler connection constraint}
%Each CZ gate needs to satisfy that the gate qubits should be mapped to physical qubits coupled by couplers.
%\subsubsection{Parallel constraint}\label{subsubsect parallel constraint}
To achieve the highest fidelity, calibration of quantum gate parameters is crucial \cite{shindi2023model}. 
In addition, we also need to calibrate the compensation pulses for the spectator couplers neighboring the gate qubits.
Initially, all the spectator qubits surrounding the gate qubits are at idle frequencies. 

\REVISE{When executing multiple CZ gates in parallel, the frequency configuration of neighboring CZ gates should ideally avoid frequency crowding. 
However, as mentioned in \Cref{subsect limit of cp}, the count of parallel situations increases exponentially with the chip size.
Frequency configuration, gate, and compensation pulse calibration for all situations are impractical. 
In other words, frequency crowding is inevitable.}

\begin{figure}[htbp]
\centerline{
\includegraphics[width=0.3\textwidth]{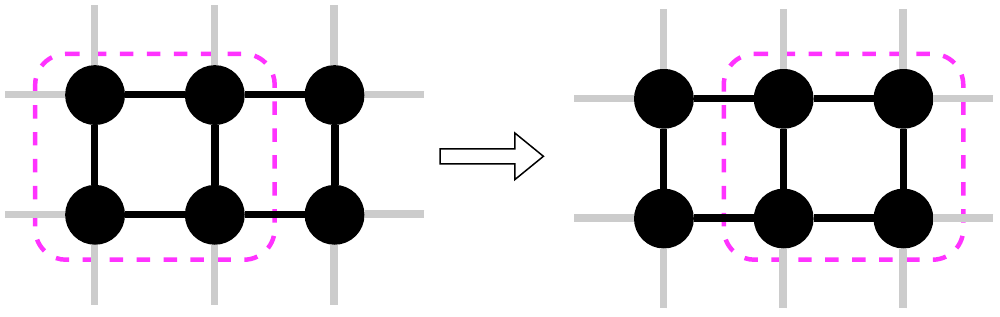}
}
\caption{
The qubits and couplers within the $m\times n, m=n=2$ window are calibrated to enable parallel execution of any CZ gates. 
By sliding and calibrating this window across the chip, any gates in $m\times n$ window at any position on the chip can be executed parallelly.}
\label{fig window opt step}
\end{figure}

An alternative approach involves calibrating every $m\times n$ ($m,n<N$) qubit window on the chip, as illustrated in \Cref{fig window opt step}. 
We perform a frequency configuration and parameter calibration considering every possible parallel scenario within the window.
For an $M \times N$ chip, the number of windows, $(M-m+1)(N-n+1)$, is of the same order of magnitude as the number of qubits. 
Configuring frequencies and calibrating gate and compensation pulse parameters for all scenarios before circuit execution is feasible.
When scheduling the execution of CZ gates, 
they must be mapped to physical qubits within the same window or non-adjacent windows to avoid unintended frequency crowding.

Building upon this concept, the parallel constraint arises from the maximum window size of $m\times n$. 
If a set of gates $\bm{g}$ temporally overlaps, expressed as:
%\begin{small}
\begin{equation}
\forall g_i,g_j\in\bm{g}, (g_i.t,g_i.t+t_{g_i})\cap(g_j.t,g_j.t+t_{g_j})\neq\emptyset.\label{eq overlap cons}
\end{equation}
%\end{small}
Gate qubits should be mapped to physical qubits within non-adjacent windows.

% \begin{figure}[htbp]
% \centerline{
% \includegraphics[width=0.25\textwidth]{window.pdf}
% }
% \caption{
% The red edges are spectator couplers that require compensation pulses.
% Assuming the window size is $2\times 2$, the CZ gates mapped to the left window violate the constraint, with $d_g=3$ exceeding the limit of 2.
% On the other hand, the CZ gates mapped to the right window satisfy the constraint, with a diameter of $d_g=2$, which is less than or equal to 2.}
% \label{fig window}
% \end{figure}
Suppose the maximum allowed window size is $m\times n$, with the graph diameter of the window being $m+n-2$.
Given a mapping $\pi$ and a list of pending gates $\bm{g}$, we define a subset $\bm{Q}_g=\{\pi(q)|\forall g\in\bm{g},q\in g.q\}$, and obtain an algorithm subgraph $\mathcal{G}_g$ of $\mathcal{G}$ containing only $\bm{Q}_g$.
For a maximum window with size $m\times n$, the constraint can be expressed as follows:
\begin{equation}\label{eq parallel cons}
    \max(d_g)\leqslant m+n-2,
\end{equation}
where $d_g$ represents the graph diameter of all connected subgraphs of $\mathcal{G}_g$.
\REVISE{These windows can locally mitigate crosstalk. 
The window size depends on the frequency configuration and pulse parameter calibration scale that the chip control system can simultaneously achieve, 
as described in \Cref{sect compensation pulse}. 
It determines the maximum number of CZ gates that the mapping and scheduling approach can tolerate with frequency crowding.
In the following mapping and scheduling approach, we will explain how to leverage these local windows to achieve crosstalk and decoherence suppression across the entire chip.
}

\subsection{Mapping algorithm}
\subsubsection{Key design}
Our primary design strategy involves a delay in gate execution when the parallel constraint \Cref{eq parallel cons} is violated. 
This delay results in an extension of the execution time, denoted as $t_\text{end}$. 
We define a score function to evaluate the quality of the mapping, as shown in \Cref{eq score}:
%\begin{small}
\begin{align}
\text{score}=\frac{|g_\text{exc}|-3|s|}{t_\text{end}}.\label{eq score}
\end{align}
%\end{small}
Here, the numerator serves as a reward, where $|g_\text{exc}|$ denotes the number of gates that can be executed, and $|s|$ represents the number of swap gates inserted, encouraging more gate executions and fewer swap gates. 
The denominator, representing the execution time $t_\text{end}$, serves as a penalty, discouraging mappings susceptible to significant crosstalk and longer execution times.

\begin{figure}[htbp]
\centering
\begin{minipage}[t]{0.29\linewidth}
\centering
\includegraphics[width=0.8\textwidth]{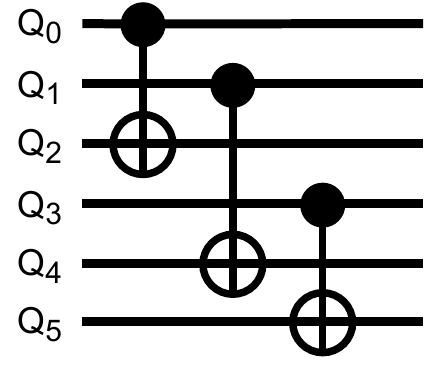}\\
(a)
\end{minipage}
\begin{minipage}[t]{0.3\linewidth}
\centering
\includegraphics[width=0.9\textwidth]{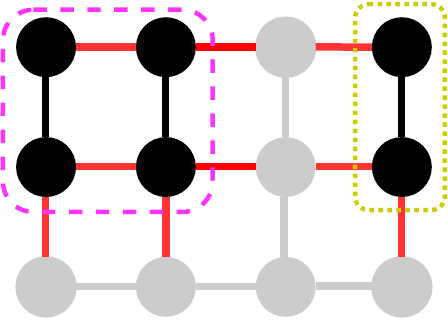}\\
(b)
\end{minipage}
\begin{minipage}[t]{0.3\linewidth}
\centering
\includegraphics[width=0.9\textwidth]{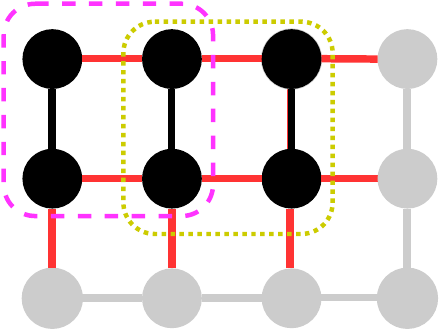}\\
(c)
\end{minipage}
\caption{
(a) The circuit comprises three CZ gates. (b) and (c) depict two distinct mappings on the chip. 
In (b), the mapping of CZ gates to two disjoint windows satisfies the constraint, 
while the mapping in (c) violates the constraint.}
\label{fig mapping toy scheme}
\end{figure}

Consequently, CAMEL effectively aims to minimize both decoherence and crosstalk. As illustrated in \Cref{fig mapping toy scheme}(a), three CZ gates are ready for execution. 
\Cref{fig mapping toy scheme}(b) and (c) illustrate two distinct mappings. 
Given the maximum window size of $2 \times 2$, the mapping in (b) satisfies the constraint, while the mapping in (c) does not. 
Consequently, at least one gate in (c) is delayed due to crosstalk, whereas all three gates in (b) can be executed in parallel. Our algorithm encourages the mapping of (b) over (c).

The algorithm outlined in \Cref{alg xtalk-aware mapping} details the crosstalk-aware mapping process. It starts by initializing a random mapping $\pi_0$ and an empty DAG $\mathcal{D}o$. 
Then, it iterates over the gate set $g$ in $\mathcal{D}$, utilizing the function \textbf{searchForward} to identify a subset of gates $g_\text{exc}$ for minimal noise execution. 
This function receives the current mapping $\pi_l$, DAG $\mathcal{D}$, as well as the search depth $L$ and search width $W$ as its inputs.
Gates from $g_\text{exc}$ are then transferred from $\mathcal{D}$ to $\mathcal{D}o$. 
Finally, any swap gates in $g_\text{exc}$ are applied to update the current mapping $\pi_l$.
\begin{algorithm}[htbp]
  \caption{Crosstalk-aware mapping}\label{alg xtalk-aware mapping}
  \begin{algorithmic}[1]
  \REQUIRE Coupling Graph $\mathcal{G}(\bm{Q},\bm{E})$, DAG $\mathcal{D}(\bm{q},\bm{g},\bm{e})$, search depth $L$, search width $W$
  \ENSURE new DAG $\mathcal{D}_o$ after compilation
  \STATE get the remaining gate set $\bm{g}$ in $\mathcal{D}$
  \STATE get a random initial mapping $\pi_0$
  \STATE let $\mathcal{D}_o$ be a empty DAG
  \WHILE{$\bm{g}\neq\emptyset$}
      \STATE current mapping is $\pi_l$,
      \STATE $g_\text{exc}$=searchForward($\pi_l$, $\mathcal{D}$, $L$, $W$)
      \STATE remove the gates $g_\text{exc}$ in $\mathcal{D}$
      \STATE apply the gates $g_\text{exc}$ to $\mathcal{D}_o$
      \FOR{$g$ in $g_\text{exc}$}
        \IF{$g$ is swap gate}
          \STATE apply $g$ to $\pi_l$ and get $\pi_{l+1}$
          \STATE $\pi_{l}\leftarrow\pi_{l+1}$
        \ENDIF
      \ENDFOR
      \STATE get the remaining gate set $\bm{g}$ in $\mathcal{D}$
  \ENDWHILE
\end{algorithmic}
\end{algorithm}

\subsubsection{Recursive searching forward}
The \Cref{alg fun searchForward} defines the \textbf{searchForward} function, which identifies a subset of gates for execution with minimal crosstalk, based on the current mapping $\pi_l$, input DAG $\mathcal{D}$, search depth $L$, and search width $W$.
It begins by retrieving the top layer $F$ of gates from the input DAG. An empty list $g_\text{exc}$ is initialized to store executable gates, which are selected based on coupler connection constraint and the current mapping $\pi_l$.
After adding suitable gates to $g_\text{exc}$, they are removed from the input DAG $\mathcal{D}$. If the search depth $L$ is zero, the function returns $g_\text{exc}$.
For each swap gate $s$ in $\bm{S}$, the function calculates a distance sum $d$ for the gates in $F$ under the new mapping $\pi_{l+1}$.
Afterward, the function proceeds by iterating through the first $W$ swap gates $s$ in the set $\bm{S}$, prioritized in ascending order of $d$.
Each swap gate $s$ is applied to the current mapping $\pi_l$, generating a new mapping $\pi_{l+1}$.
It then proceeds to recursively call itself with the updated parameters, including the new mapping $\pi_{l+1}$, new DAG $\mathcal{D}$, decreased search depth $L-1$, and the same search width $W$. 
This recursive call yields a list of executable gates $g_\text{exc2}$.
The function evaluates the quality of the mapping $\pi_{l+1}$ by calculating its score using the \textbf{scoreStep} function \Cref{alg fun scoreStep}.
If the score of the mapping $\pi_{l+1}$ surpasses the current maximum score (maxMapScore), it updates maxMapScore and records the list of gates $g_\text{exc,best}$ as the best choice for the current iteration.
After iterating over the first $W$ swap gates, the function returns the list of executable gates $g_\text{exc}+g_\text{exc,best}$.
\begin{algorithm}[htbp]
  \caption{Function searchForward}\label{alg fun searchForward}
  \begin{algorithmic}[1]
  \REQUIRE $\pi_l$, $\mathcal{D}$, $L$, $W$
  \ENSURE executable gates $g_\text{exc}$
  \STATE get the top layer $F$
  \STATE $g_\text{exc}=[]$
  \STATE add the $g$ in $F$ satisfied coupler connection constraint to $g_\text{exc}$
  \STATE remove $g_\text{exc}$ from $\mathcal{D}$
  \IF{$L=0$}
  \STATE return $g_\text{exc}$
  \ENDIF
  \FOR{$s$ in $\bm{S}$}
    \STATE apply $s$ to $\pi_l$ and get $\pi_{l+1}$
    \STATE $d\leftarrow\sum_{g\in F}D(\pi_{l+1}(g.q_1),\pi_{l+1}(g.q_2))$
  \ENDFOR
  \STATE maxMapScore$=-\infty$
  \FOR{the first $W$ $s$ in $\bm{S}$ in ascending order $d$}
    \STATE apply $s$ to $\pi_l$ and get $\pi_{l+1}$
    \STATE $g_\text{exc2}=$searchForward$(\pi_{l+1},\mathcal{D},L-1,W)$
    \STATE mapScore$\leftarrow$scoreStep$(\pi_l,g_\text{exc}+s+g_\text{exc2})$
    \IF{mapScore$>$maxMapScore}
      \STATE $g_\text{exc,best}\leftarrow s+g_\text{exc2}$
    \ENDIF
  \ENDFOR
  \STATE return $g_\text{exc}\leftarrow g_\text{exc}+g_\text{exc,best}$
\end{algorithmic}
\end{algorithm}

\subsubsection{Scoring strategy}
\begin{algorithm}[htbp]
  \caption{Function scoreStep}\label{alg fun scoreStep}
  \begin{algorithmic}[1]
  \REQUIRE $\pi,g_\text{exc}$
  \ENSURE score
  \STATE $Q_t\leftarrow$ a dictionary with $Q_t[Q]=0,\forall Q\in\bm{Q}$
  \STATE layers$\leftarrow[]$
  \STATE $|s|\leftarrow$ the number of gate $g\in\bm{S}$
  \FOR{$g$ in $g_\text{exc}$}
    \IF {$g\in\bm{S}$}
      \STATE apply $g$ to $\pi$ and get a new $\pi$
    \ENDIF
    \FOR{layer in layers}
      \IF{time interval $(Q_t[\pi(g.q)],Q_t[\pi(g.q)]+t_g)$ has overlap with the gates in layer}
        \STATE $\bm{Q}_g\leftarrow\{\pi(g_l.q)|\forall g_l\in\text{layer}\cup g\}$ and 
          induce $\mathcal{G}_g$ from $\bm{Q}_g$ and $\mathcal{G}$
        \IF{there is not connected subgraph of $\mathcal{G}_g$ violate the parallel constraint}
          \STATE $Q_t[\pi(g.q)]\leftarrow Q_t[\pi(g.q)]+t_g$
          \STATE layer.append($g$)
        \ELSE
          \STATE $t\leftarrow$ the maximum $Q_t[\pi(g_l.q)]$ for gate qubits $g_l.q$ in layer
          \STATE $Q_t[\pi(g.q)]\leftarrow t$
        \ENDIF
      \ENDIF
    \ENDFOR
    \IF{there is no layer for $g$}
      \STATE layers.append([$g$])
      \STATE $t\leftarrow$ the maximum time in $Q_t$
      \STATE $Q_t[\pi(g.q)]\leftarrow t+t_g$
    \ENDIF
  \ENDFOR
  \STATE $t_\text{end}\leftarrow$ the maximum time in $Q_t$
  \STATE score$=\frac{|g_\text{exc}|-3|s|}{t_\text{end}}$
  \STATE return score
\end{algorithmic}
\end{algorithm}

The purpose of function \textbf{scoreStep} in \Cref{alg fun scoreStep} is to evaluate the score of a current mapping and a list of executable gates. 
It initializes a dictionary $Q_t$ to track the time of each physical qubit, and creates a list layers to accommodate gates that can execute parallelly. 
Additionally, it sets $|s|$ to store the count of swap gates in the list of executable gates $g_\text{exc}$. The algorithm then iterates over the gates $g$ in $g_\text{exc}$, applying swap gates in $g_\text{exc}$ to update the map. 
For each gate $g$, the algorithm assesses whether it overlaps with gates in the current layer. 
If it satisfies \Cref{eq overlap cons} and \Cref{eq parallel cons} alongside the gates in the layer, $g$ is placed in the layer; otherwise, it is delayed. Additionally, if $g$ does not overlap with existing gates, a new layer is created.
After processing all gates, the algorithm calculates the mapping score \Cref{eq score} based on the number of executable gates, swap gates, and execution time.

\subsubsection{Complexity analysis}
The complexity of \Cref{alg fun scoreStep} depends on the number of iterations in the gate set $|\bm{g}|$ and the number of layers $L$, resulting in a time complexity of $O(|\bm{g}|L)$. 
The most resource-intensive operation occurs in the recursive call of \Cref{alg fun searchForward}. Here, the algorithm makes a maximum of $W$ recursive calls, each with reduced depth $L-1$. 
Thus, the time complexity is described by the recurrence relation:
\begin{equation}\label{eq complexity}
\begin{aligned}
T(L,W)&=WT(L-1,W)+O(W|\bm{g}|L),\\
&=W^LT(0,W)+O\left(\sum_{l=0}^{L-1}W^l|\bm{g}|L\right),\\
&=O\left(W^L(|\bm{g}|L+1)\right).
\end{aligned}
\end{equation}
Here, $|\bm{g}|$ represents the number of gates in $\mathcal{D}$. Since \Cref{alg xtalk-aware mapping} calls the \textbf{searchForward} function at most $|\bm{g}|$ times, once for each gate in the original DAG, the time complexity can be expressed as $O\left(|\bm{g}|W^L(|\bm{g}|L+1)\right)=O\left(W^L|\bm{g}|^2L\right)$. Notably, this complexity is polynomial with respect to the circuit scale $|\bm{g}|$.

\subsection{Gate scheduling algorithm}
As the mapping algorithm is heuristic, it cannot entirely eliminate the gate time delay problem resulting from frequency crowding. 
Consequently, finding an optimal way to select a gate execution order that minimizes the circuit execution time becomes necessary.
\subsubsection{Barrier inserting}
\begin{figure}[htbp]
\centering
\begin{minipage}[t]{0.4\linewidth}
\centering
\includegraphics[width=\textwidth]{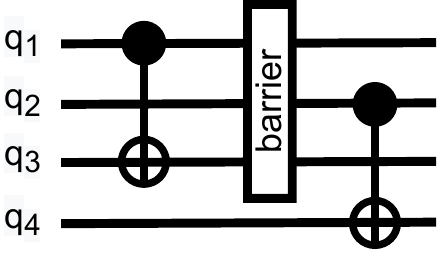}\\
(a)
\end{minipage}
\begin{minipage}[t]{0.55\linewidth}
\centering
\includegraphics[width=\textwidth]{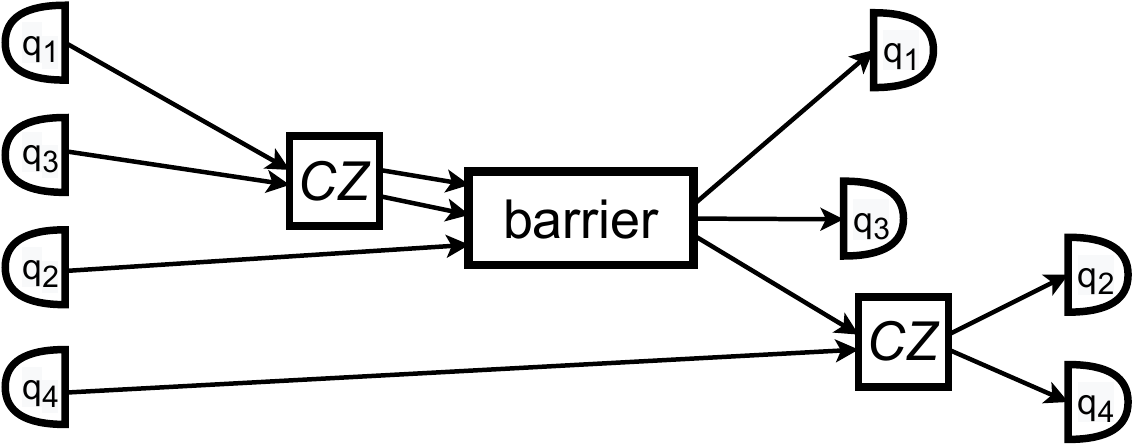}\\
(b)
\end{minipage}
\caption{
(a) A barrier was inserted between these CZ gates.
(b) Gate order is changed as follows: $\text{CZ}_1<$barrier$<\text{CZ}_2$.}
\label{fig barrier}
\end{figure}

\begin{algorithm}
  \caption{Gate scheduling algorithm}\label{alg scheduling}
  \begin{algorithmic}[1]
  \REQUIRE DAG $\mathcal{D}(\bm{q},\bm{g},\bm{e})$, Coupling Graph $\mathcal{G}(\bm{Q},\bm{E})$
  \ENSURE gTime
  \STATE gTime=extractGateTime$(\mathcal{D})$
  \STATE layers$\leftarrow[]$
  \FOR{$g$ in $\bm{g}$}
    \FOR{layer in layers}
      \IF{$\exists g_l\in$layer has execution time overlap with gTime$[g]$}
        \STATE layer.append($g$)
      \ENDIF
    \ENDFOR
    \IF{there is no layer for $g$}
      \STATE layers.append$([g])$
    \ENDIF
  \ENDFOR
  \STATE partitions=generatePartition(layers,$\mathcal{G}$)
  \FOR{(partition, layer) in (partitions, layers)}
    \STATE add barrier to gates in layer according to partition
  \ENDFOR
  \STATE gTime=extractGateTime$(\mathcal{D})$
  \STATE return gTime
\end{algorithmic}
\end{algorithm}
The \Cref{alg scheduling} takes the DAG $\mathcal{D}(\bm{q},\bm{g},\bm{e})$ and the coupling graph $\mathcal{G}(\bm{Q},\bm{E})$ as inputs. It produces the gate time for each gate in the circuit.
Initially, the algorithm utilizes the function \textbf{extractGateTime} to assign gate times based on gate durations and circuit dependencies. Subsequently, it arranges the gates into layers, where gates within each layer overlap in time.
Next, the algorithm employs the function \textbf{generatePartition} to divide the layers into sub-layers, ensuring that gates within each sub-layer can be executed parallelly without violating \Cref{eq parallel cons}. To ensure sequential execution of gates across different sub-layers, barriers are inserted between gates among each layer.
In the provided example shown in \Cref{fig barrier}, due to the presence of barriers, the execution of the second CZ gate involving qubits $q_2$ and $q_4$ is delayed, as it relies on the barrier involving qubits $q_1$, $q_2$, and $q_3$.

\subsubsection{Maximum-independent set partition}
To depict the crosstalk relationship between parallel gates, we first introduce the crosstalk graph $\mathcal{T}(\bm{g},\bm{X})$, 
where nodes $\bm{g}$ represent CZ gates executed parallelly in the same layer. An edge $x\in\bm{X}$ connects nodes if crosstalk exists between them, defined as:
\begin{equation}\label{eq xtalk g}
\begin{aligned}
&\min(D(\pi(g_1.q),\pi(g_2.q)))=1,\\
\Rightarrow&(g_1,g_2)\in\bm{X}.
\end{aligned}
\end{equation}
\Cref{eq xtalk g} implies that when the logical qubits of parallel CZ gates are mapped to physical qubits on the chip with a minimum distance of 1, crosstalk between the gates occurs.

In \Cref{alg generatePartition}, the function \textbf{generatePartition} iterates over each layer to seek out the maximum-independent sets i.e., subsets of crosstalk-free gates within $\mathcal{T}$.
Firstly, mapping the CZ gates in this layer to the chip. Initialize a window list $l_{w_i}$ for each window $w_i$ containing pending CZ gate in layer. Iterate through each window $w_j$ on the chip,
if $w_j$ contains qubits with a minimum distance greater than 2 from all qubits in $l_{w_i}$, then $w_j$ is added to $l_{w_i}$. This step aims to identify the largest set of non-adjacent windows that can be executed parallelly without crosstalk. 
Next, select the $l_{w}$ with the max coverage of pending gates, forming a covering set capable of executing the max number of CZ gates simultaneously.

Following this, the coupler edges between CZ gates covered by windows are removed from the algorithm subgraph $\mathcal{G}_g$. 
This step indicates the mitigation of crosstalk between CZ gates covered by windows through compensation pulses. 
Based on the resulting $\mathcal{G}_g$ after edge deletion, we obtain $\mathcal{T}$ according to \Cref{eq xtalk g}.
Utilizing Python library Networkx \cite{hagberg2020networkx}, we apply the \textbf{maxIndependentSet} function to find solutions to maximum-independent set problem of $\mathcal{T}$ in polynomial time. 

\begin{figure}[htbp]
\centering
\includegraphics[width=0.48\textwidth]{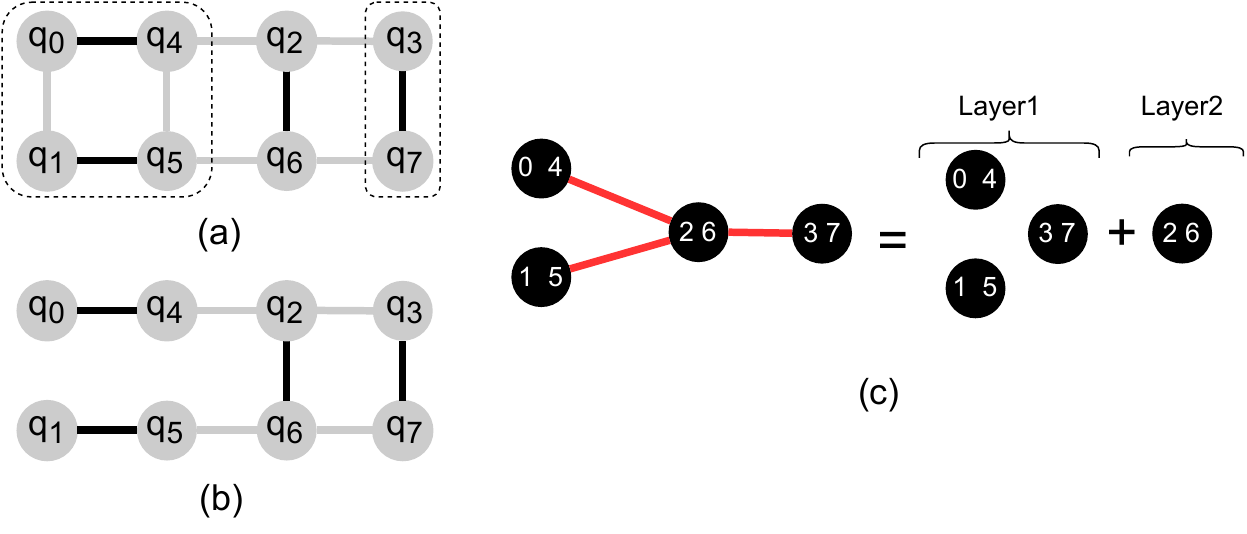}\\
\caption{
(a) The mapping of CZ gates on chip, with black edges indicating activated couplers.
Qubits $q_0,q_4$ and $q_1,q_5$ are covered by window $w_1$, and $q_3,q_7$ are covered by another non-adjacent window $w_2$.
(b) The edges within these windows are deleted.
(c) The crosstalk graph consists of nodes representing parallel gates from (a). This graph is partitioned into two maximum-independent sets.}
\label{fig max ind set}
\end{figure}

\begin{figure*}
\centering
\includegraphics[width=\textwidth]{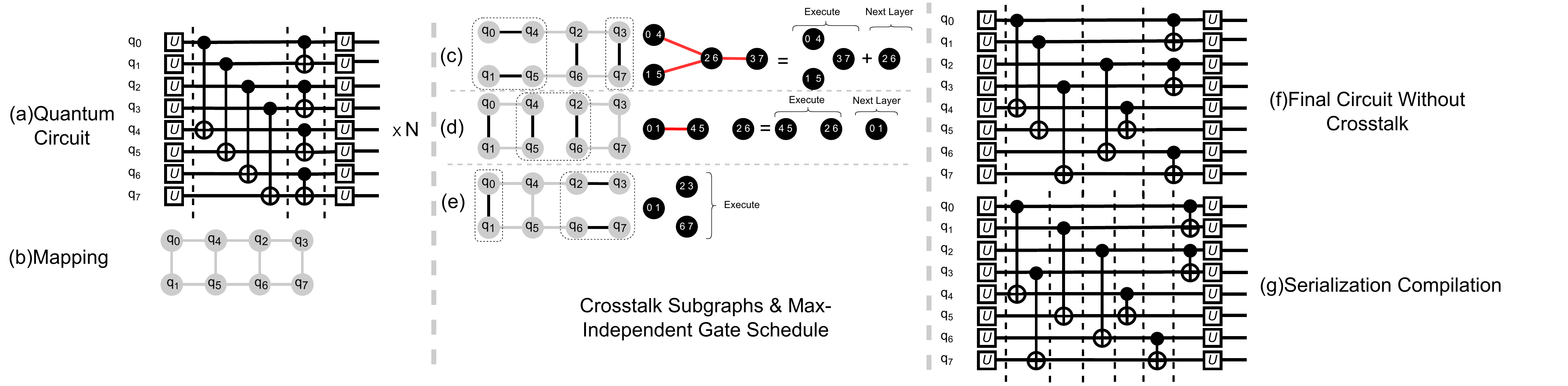}
\caption{
The schematic of the compilation process is illustrated assuming our chip has a maximum calibration window size of $2\times 2$. (a) depicts a segment of the quantum circuit from the VQE algorithm \cite{adedoyin2018quantum}. 
In (b), a mapping to qubits on a $2\times 4$ quantum chip is illustrated.
(c-e) demonstrate the parallel execution on the chip of the top layer gates of the quantum circuit according to this mapping, accompanied by the corresponding crosstalk subgraph. 
In (c), it's evident that executing $g_{q_0,q_4}g_{q_1,q_5}g_{q_2,q_6}g_{q_3,q_7}$ parallelly violates the parallel constraint.
On the other hand, our gate scheduling approach finds the maximum parallel execution of the first three gates: $g_{q_0,q_4}$, $g_{q_1,q_5}$, and $g_{q_3,q_7}$.
(c-e) illustrate gate scheduling steps based on the maximum independent set. (f) presents the compiled circuit, completing execution in only three layers. 
Conversely, executing all crosstalk gates serially, as depicted in figure (g), would necessitate six layers, increasing decoherence.}
\label{fig flow}
\end{figure*}

\begin{algorithm}
  \caption{Function generatePartition}\label{alg generatePartition}
  \begin{algorithmic}[1]
  \REQUIRE layers, Coupling Graph $\mathcal{G}(\bm{Q},\bm{E})$
  \ENSURE partitions is a dictionary whith partitions[$g]=i$
  which means that $g$ in the $i^{\text{th}}$ partition
  \FOR{layer in layers}
    \STATE $\bm{l}=[]$
    \FOR{window $w_i$ include $g\in$ layer on the chip}
      \STATE $l_{w_i}\leftarrow[w_i]$
        \FOR{window $w_j\notin l_{w_i}$ on the chip}
          \IF{all qubits in $w_j$ with a distance $>2$ to all qubits in $l_{w_i}$.}
            \STATE $l_{w_i}$.append($w_j$)
          \ENDIF
        \ENDFOR
        \STATE $\bm{l}$.append($l_{w_i}$)
    \ENDFOR
    \STATE Select window list $l_{w}$ in $\bm{l}$ which covers maximum qubit set of layer
    \STATE Remove coupler edges of $\mathcal{G}_g$ between CZ gates covered by $l_{w}$
    \STATE Calculate $\mathcal{T}$ based on $\mathcal{G}_g$ using \Cref{eq xtalk g}
    \STATE independentSets $\gets$ maxIndependentSet($\mathcal{T}$)
    \STATE Divide layer according to independentSets
  \ENDFOR
\end{algorithmic}
\end{algorithm}

\Cref{fig max ind set} serves as an example.
It's found that the window $w_1$ corresponding to $q_0,q_4$ and $q_1,q_5$ and another non-adjacent window $w_2$ corresponding to $q_3,q_7$, cover the maximum of CZ gates. Consequently, the edges between the two gates within $w_1$ are removed, resulting in the crosstalk graph $\mathcal{T}$. Employing the \textbf{maxIndependentSet} function, $\mathcal{T}$ is divided into two independent subgraphs. This indicates that the four CZ gates will be split into two steps: the first step executes the gates between $[q_0,q_4];[q_1,q_5];[q_3,q_7]$, while the second step executes the gate between $[q_2,q_6]$.

\subsubsection{Complexity analysis}
The complexity of \Cref{alg generatePartition} is $O(|\bm{g}|(|\mathcal{G}|+N^2))$. Here, $|\mathcal{G}|$ represents the complexity of finding a maximum-independent set, and $N$ denotes the number of qubits on the chip. 
$N^2$ indicates the complexity of identifying the maximum cover of pending gates. The \textbf{maxIndependentSet} function in Networkx has a complexity of $O(|\mathcal{G}|)=O(|\bm{g}|/(\log^2|\bm{g}|))$ \cite{hagberg2020networkx}, 
where $|\bm{g}|$ corresponds to the node number of $\mathcal{G}$, which is at most the number of gates. \Cref{alg generatePartition} calls the \textbf{maxIndependentSet} function at most $|\bm{g}|$ times. 
Considering that \Cref{alg scheduling} invokes \Cref{alg generatePartition} once and the complexities of the loop and \textbf{extractGateTime} are $|\bm{g}|^2$, 
the overall complexity amounts to $O(|\bm{g}|^2+|\bm{g}|(|\mathcal{G}|+N^2))=O(|\bm{g}|^2(1/(\log^2|\bm{g}|)+1)+|\bm{g}|N^2)=O(|\bm{g}|^2+|\bm{g}|N^2)$.

\section{Evaluation}

\subsection{Baselines}
In this section, we evaluate the CAMEL algorithm and compare it with several baselines as follows:
\begin{itemize}
\item \textbf{Crosstalk-agnostic compilation (N)}: This approach relies on fixed idle and interaction frequencies without optimization for crosstalk mitigation~\cite{zhang2021time, venturelli2017temporal, zulehner2018efficient, li2019tackling}. It employs a crosstalk-agnostic qubit mapper and a tiling gate scheduler. We adopt the Sabre approach as a representation of crosstalk-agnostic compilation~\cite{li2019tackling}.
\item \textbf{Serialization compilation (S)}: This approach utilizes fixed idle and interaction frequencies without optimization for crosstalk mitigation~\cite{murali2020software, Hua2022CQCAC}. It employs a crosstalk-aware gate scheduler that serializes parallel CZ gates. We adopt the approach proposed by Murali et al.~\cite{murali2020software} to represent serialization compilation.
\item \textbf{Static frequency-aware compilation (SF)}: In this approach, idle and interaction frequencies are fixed and optimized for crosstalk mitigation \cite{klimov2020snake}. It employs a crosstalk-aware gate scheduler. We utilize the snake optimizer as a representation of static frequency-aware compilation.
\item \textbf{Dynamic frequency-aware compilation (DF)}: Idle frequencies remain fixed while interaction frequencies are dynamically optimized for each quantum circuit. Additionally, this approach utilizes a crosstalk-aware gate scheduler. We adopt Ding's approach~\cite{Ding2020SystematicCM} as a baseline of this type.
\item \textbf{CAMEL (this paper)}: CAMEL utilizes a crosstalk-aware mapper and gate scheduler, which employs fixed optimized idle and interaction frequencies, along with compensation pulse to mitigate crosstalk.
\end{itemize}

\subsection{Architectural features}
% We utilize a chip architecture comprising a 2D grid of $N\times N$ frequency-tunable qubits and couplers.
% The qubits operate within the frequency range of $\omega_q\in(4, 5)\text{GHz}$, while the couplers span $\omega_c\in(5, 7)\text{GHz}$.
% The anharmonicity values for the couplers and qubits are approximately $\eta_c\approx-100\text{MHz}$ and $\eta_q\approx-200\text{MHz}$.
% The coupling strengths $g_{ic}\simeq100\text{MHz}$ and $g_{12}\simeq10\text{MHz}$. 
% Each frequency-tunable qubit is connected by frequency-tunable couplers. The decoherence time $T_i$ are modeled based on \cite{ithier2005decoherence}.
% Furthermore, both initial and measurement errors are set within a range of $0.01\pm 0.001$.
% These values are obtained from experimental data in the literature \cite{kjaergaard2020programming}.
We consider a chip architecture consisting of a 2D grid of $N \times N$ frequency-tunable qubits and couplers. 
The qubits operate within a frequency range of $\omega_q \in (4, 5)\text{GHz}$, while the couplers span $\omega_c \in (5, 7)\text{GHz}$. 
The anharmonicity values for the couplers and qubits are approximately $\eta_c \approx -100\text{MHz}$ and $\eta_q \approx -200\text{MHz}$, respectively. 
The coupling strengths are $g_{ic} \simeq 100\text{MHz}$ and $g_{12} \simeq 10\text{MHz}$. 
Each frequency-tunable qubit is connected via frequency-tunable couplers. The decoherence times $T_i$ are modeled based on \cite{ithier2005decoherence}. 
Additionally, both initial and measurement errors are set within a range of $0.01 \pm 0.001$. 
These values are obtained from experimental data reported in the literature \cite{kjaergaard2020programming}.

\begin{figure*}
\centering
\begin{minipage}[t]{0.7\linewidth}
\centering
\includegraphics[width=\textwidth]{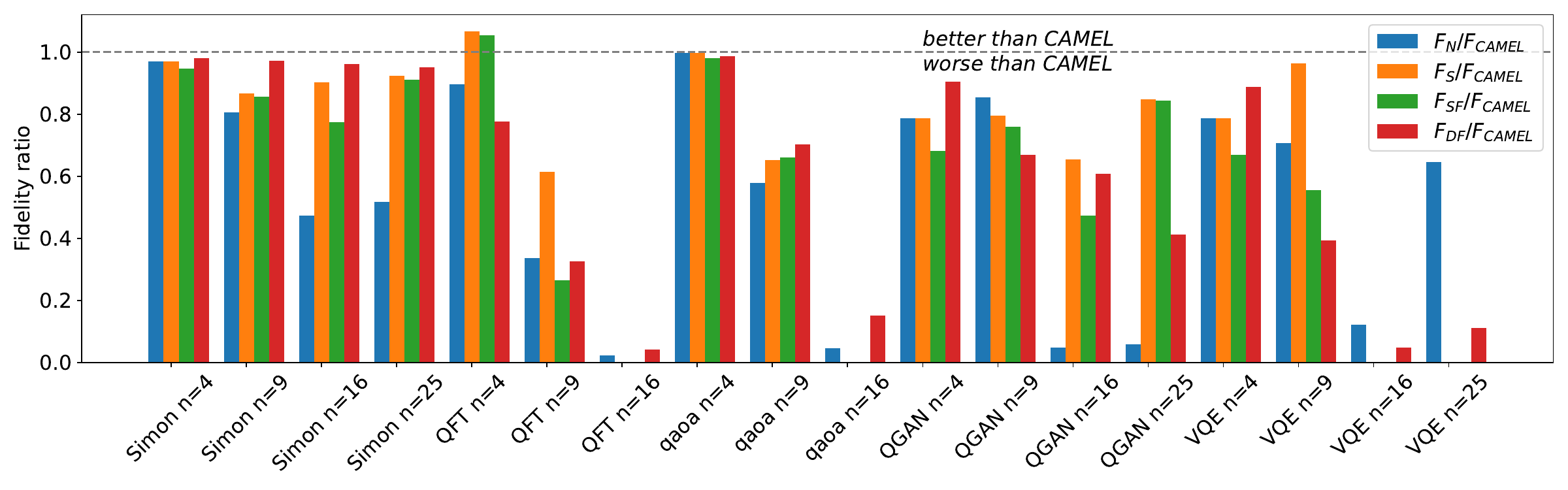}\\
(a)
\end{minipage}
\begin{minipage}[t]{0.29\linewidth}
\centering
\includegraphics[width=\textwidth]{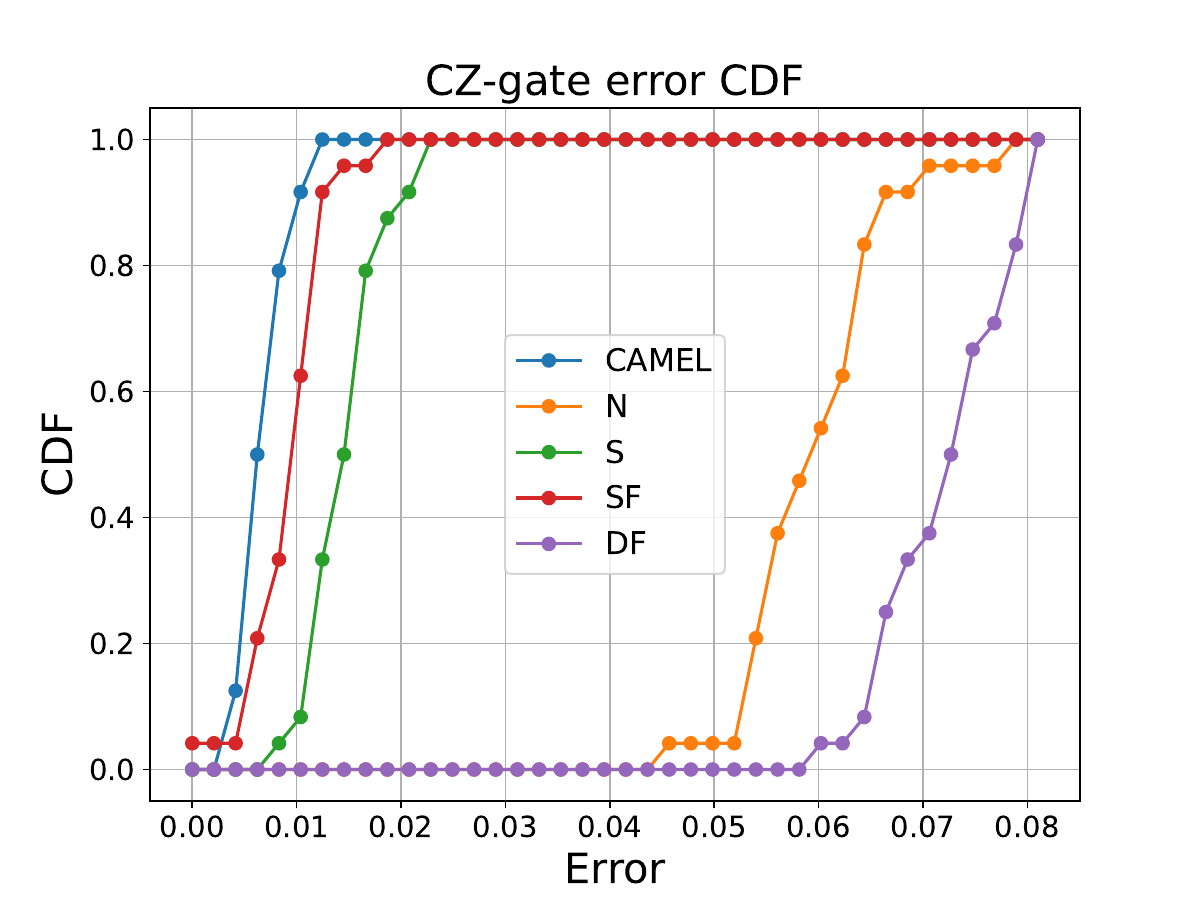}
(b)
\end{minipage}
\caption{
(a) The fidelity of the Simon algorithm, QFT algorithm, QAOA algorithm, QGAN algorithm, and VQE algorithm after being compiled with different baselines. 
We have computed the fidelity ratios between all baselines and CAMEL approach. The gray dashed line represents a ratio of one. Bars above the dashed line indicate better performance than CAMEL, 
while bars below the dashed line indicate worse performance.
Some results for 16 and 25-qubit QFT, QAOA, and VQE algorithms are absent due to the excessive simulation time. It can be observed that CAMEL approach consistently maintains high fidelity.
(b) The XEB experiment of the approaches. 
Through parallel XEB circuits, the error of each two-qubit gate on the chip can be obtained, resulting in a cumulative distribution function (CDF) plot.
CAMEL have the lowest error distribution, the second is static-frequency aware compilation, and then, 
serialization compilation, crosstalk-agnostic compilation and the worst one is dynamicfrequency-aware compilation.
}
\label{fig fidelity}
\end{figure*}

\subsection{Benchmarks}
% We evaluate the performance of our algorithm using NISQ benchmarks from \cite{adedoyin2018quantum}, which are key applications for near-term quantum devices. 
% Additionally, we utilize cross entropy benchmarking (XEB) circuits \cite{arute2019quantum} to demonstrate the effect of crosstalk on gate fidelity.
% \begin{itemize}
% \item \textbf{Simple quantum algorithms}: Including Simon's algorithm and quantum fourier transformation (QFT) \cite{nielsen2002quantum}.
% \item \textbf{Quantum optimization algorithm}: Quantum approximate optimization algorithm (QAOA) \cite{choi2019tutorial} applied to MAX-CUT on an Erdos-Renyi random graph.
% \item \textbf{Variational quantum algorithm}: Employing variational quantum eigensolver (VQE) to determine the ground state energy of molecules.
% \item \textbf{Quantum machine learning algorithm}: Utilizing quantum generative adversarial networks (QGAN).
% \item \textbf{Cross entropy benchmarking}: Employing XEB circuits with 16 qubits and 200 cycles.
% \end{itemize}
% The circuits mentioned in the evaluation have varying numbers of qubits: 4, 9, 16, and 25.
We evaluate the performance of our algorithm using NISQ benchmarks from \cite{adedoyin2018quantum}, which represent key applications for near-term quantum devices. 
In addition, we use cross entropy benchmarking (XEB) circuits \cite{arute2019quantum} to demonstrate the effect of crosstalk on gate fidelity.
\begin{itemize}
    \item \textbf{Simple quantum algorithms}: Including Simon's algorithm and the Quantum Fourier Transform (QFT).
    \item \textbf{Quantum optimization algorithm}: Quantum Approximate Optimization Algorithm (QAOA) \cite{choi2019tutorial} applied to MAX-CUT on an Erdős-Rényi random graph.
    \item \textbf{Variational quantum algorithm}: Using the Variational Quantum Eigensolver (VQE) to determine the ground state energy of molecules.
    \item \textbf{Quantum machine learning algorithm}: Utilizing Quantum Generative Adversarial Networks (QGAN).
    \item \textbf{Cross entropy benchmarking}: Using XEB circuits with 16 qubits and 200 cycles.
\end{itemize}
\begin{figure*}
\centering
\begin{minipage}[t]{0.3\linewidth}
\centering
\includegraphics[width=\textwidth]{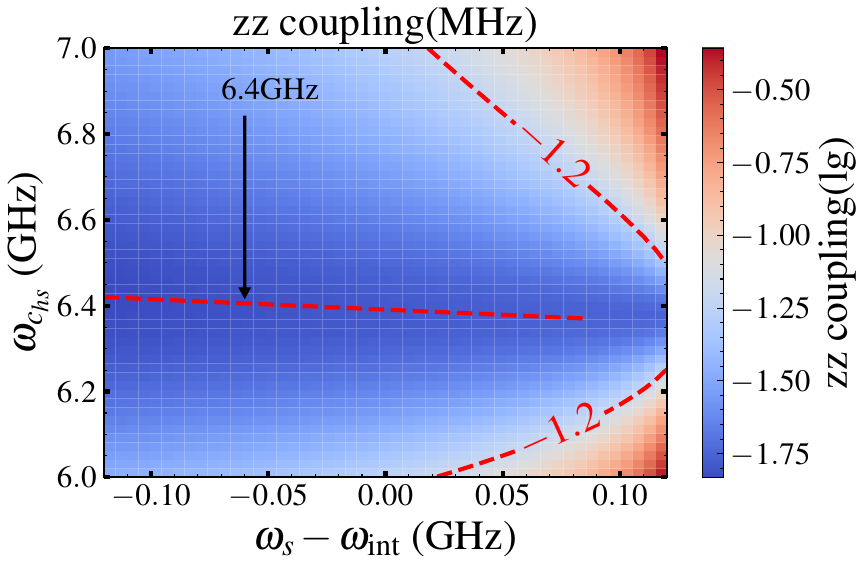}\\
(a)
\end{minipage}
\begin{minipage}[t]{0.3\linewidth}
\centering
\includegraphics[width=\textwidth]{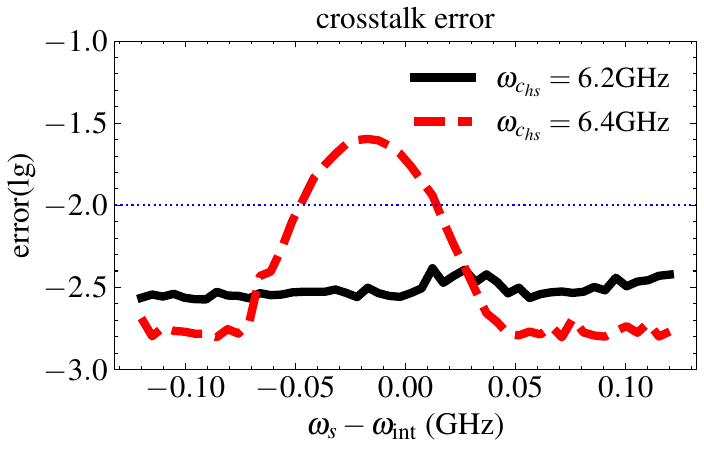}\\
(b)
\end{minipage}
\begin{minipage}[t]{0.3\linewidth}
\centering
\includegraphics[width=\textwidth]{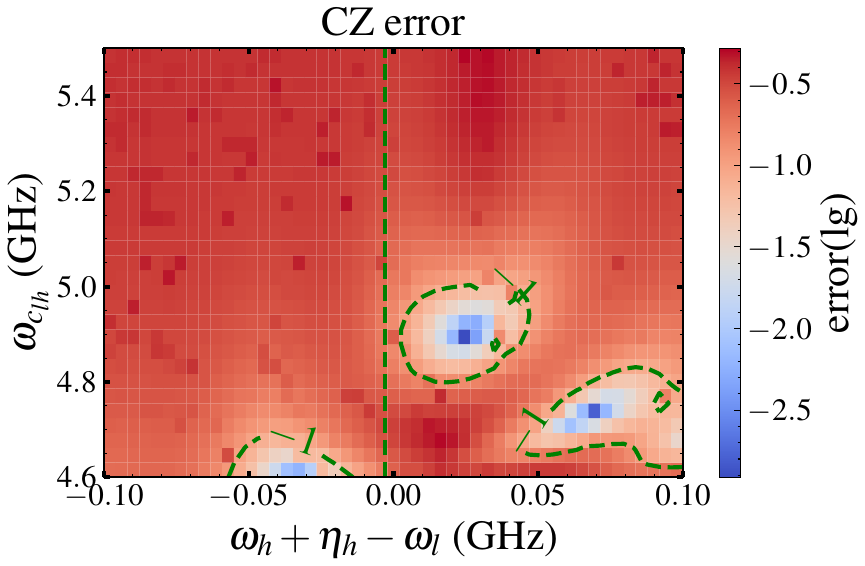}\\
(c)
\end{minipage}
\caption{
(a) The graph illustrates the minimum ZZ-coupling occurring around $\omega_{c_{hs}}=6.4$ GHz, with spectator qubit frequency situated around $\omega_\text{int}$.
(b) During CZ gate execution, if we don't adjust the coupler frequency to 6.2 GHz when $\omega_s$ and $\omega_\text{int}$ collide, the error will increase to $10^{-2}$.
(c) Let $\omega_{l}=\omega_{h}+\eta_{h}=\omega_\text{int}$ be the interaction frequency.
The actual interaction frequency deviates from the initially set interaction frequency. Under the initial interaction frequency, the error does not drop below $10^{-1}$.
}
\label{fig result}
\end{figure*}

\subsection{Software implementation}
Quantum gates and circuits are simulated using Qutip \cite{johansson2012qutip}.
The graph algorithm is implemented using Networkx \cite{hagberg2020networkx}.
\REVISE{Simulating a quantum circuit Hamiltonian at the pulse level is highly unscalable. 
To incorporate crosstalk effects into the simulation, we examine \Cref{eq h three}, 
which represents the population swap from gate state $\ket{020}$ to population swap state $\ket{110}$. 
The coupling strength $g_\text{xtalk}$ can be determined from the qubit frequency configuration. 
Apply a coordinate transformation $U=\exp(\frac{i}{\hbar} H_0)$, and we obtain:
\begin{equation}
\begin{aligned}
  H&=UHU^\dagger-i\hbar\dot{U}U^\dagger\\
  &=\begin{pmatrix}
  0&g_\text{xtalk}\exp\frac{i\Delta_{gs}t}{\hbar}\\
  g_\text{xtalk}\exp\frac{\Delta_{gs}t}{i\hbar}&0
  \end{pmatrix},
\end{aligned}\label{eq UH}
\end{equation}
where $\Delta_{gs}=E_{020}-E_{110}$. When the frequency crowding occurs, $\Delta_{gs}\approx 0$.
Thus \Cref{eq UH} can be rewritten as:
\begin{equation}\label{eq h two}
H/\hbar=
\begin{pmatrix}
0&g_\text{xtalk}\\
g_\text{xtalk}&0
\end{pmatrix}.
\end{equation}
\Cref{eq h two} can be solved as a unitary transformation between the states $\ket{110}$ and $\ket{020}$.
\begin{equation}\label{eq U two}
\begin{pmatrix}
\cos g_\text{xtalk}t_g&-i\sin g_\text{xtalk}t_g\\
-i\sin g_\text{xtalk}t_g&\cos g_\text{xtalk}t_g
\end{pmatrix},
\end{equation}
which means that during the gate execution time $t_g$, the population swap from the gate state $\ket{020}$ to the population swap state $\ket{110}$ can be modeled as a unitary transformation, 
i.e., a quantum gate.
Firstly, we calculate $|E_{020} - E_{110}|$ based on the frequency configuration to identify whether $\ket{020}$ experiences population swap. 
Secondly, we determine $g_\text{xtalk}$ according to frequency configuration.
After executing the two-qubit gate on $Q_1$ and $Q_2$, we apply the quantum gate \Cref{eq U two} to $Q_1$, $Q_2$, and $Q_s$.
The simulation is still performed at the gate level, avoiding the direct Hamiltonian simulation of pulses. 
}

\begin{figure*}
\centering
\begin{minipage}[t]{0.7\linewidth}
\centering
\includegraphics[width=\textwidth]{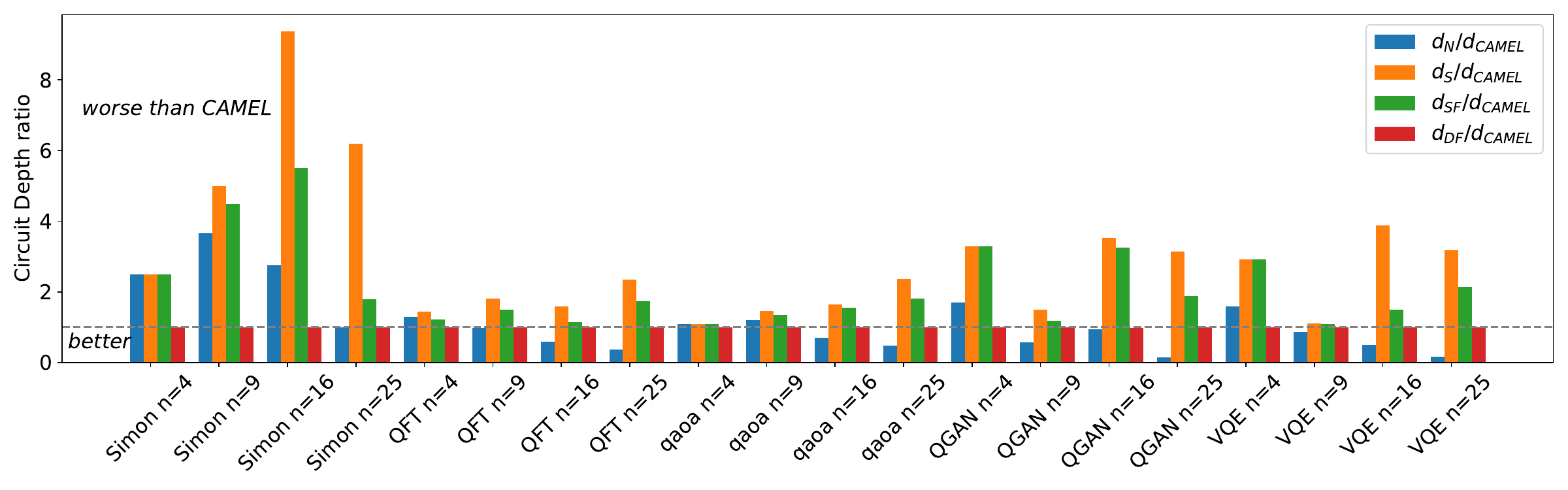}\\
(a)
\end{minipage}
\begin{minipage}[t]{0.29\linewidth}
\centering
\includegraphics[width=\textwidth]{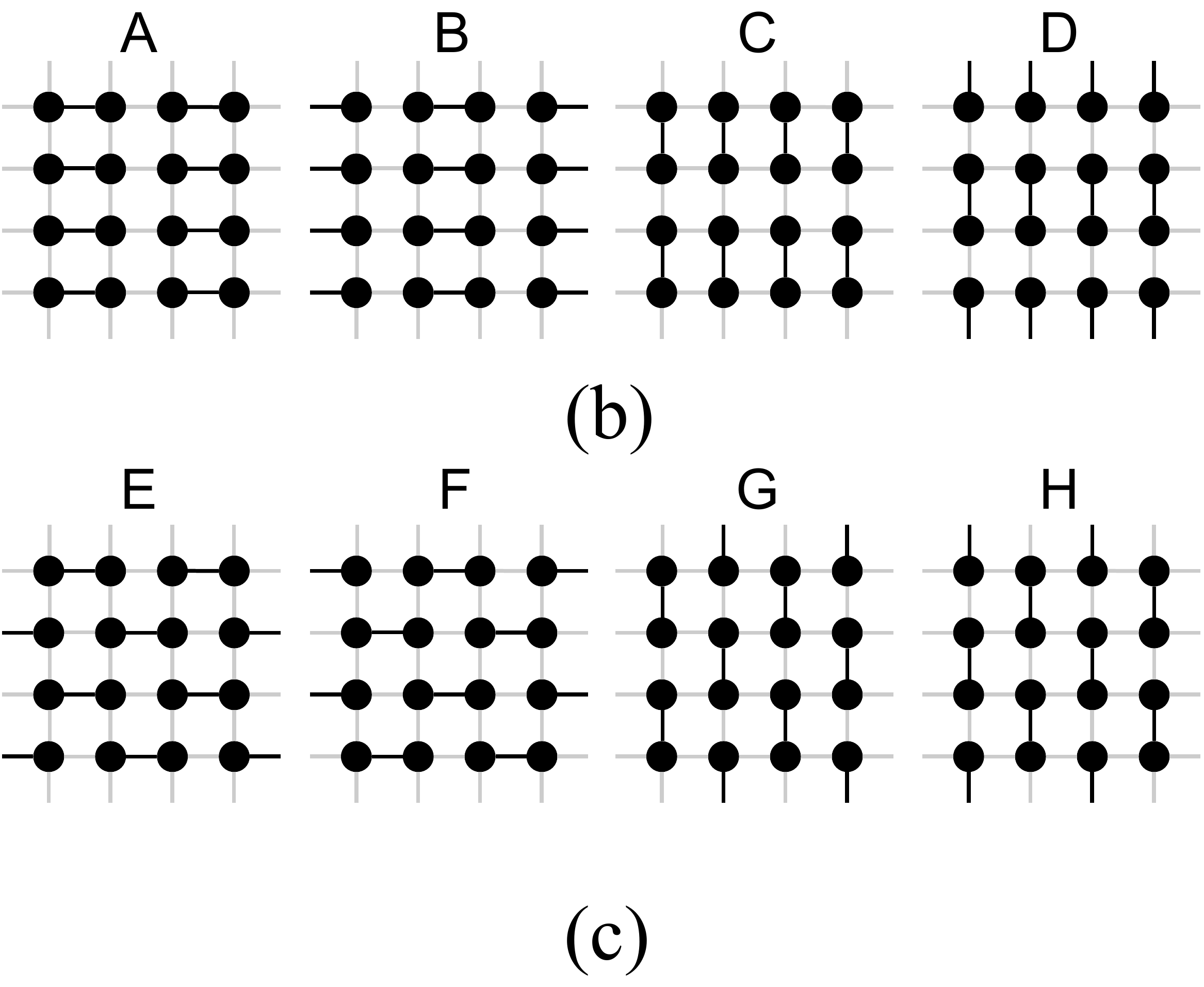}\\
\end{minipage}
\caption{
(a) The depth ratio of benchmark circuits compiled with different baselines. The gray dashed line represents a ratio of one. 
As depicted, the time durations corresponding to the serialization and static frequency baselines are generally longer than those of CAMEL.
(b-c) Coupler activation patterns. Coupler activation pattern determines which qubits are allowed to execute CZ gate
simultaneously in a cycle. ABCD patterns in (b) are exclusive from the EFGH in (c).
}
\label{fig compilation}
\end{figure*}

\subsection{Results}
\Cref{fig fidelity}(a) is the comparation of compiled circuits fidelity. Each bar is the fidelity ratio between compared baseline and CAMEL approach, the higher the better.
The gray dashed line represented ratio one.
Based on \Cref{fig fidelity}(a), the fidelity of CAMEL is generally higher than that of other approaches. In the XEB experiment \Cref{fig fidelity}(b), CAMEL has the lowest CZ gate error distribution. 
Next, we will proceed to compare and explain each case individually.

\subsubsection{Comparison with crosstalk-agnostic compilation}
Our algorithm consistently achieves higher fidelity compared to the crosstalk-agnostic compilation baseline due to its consideration of crosstalk. 
The crosstalk-agnostic approach overlooks crosstalk, resulting in frequency collisions and significantly lower gate fidelity. 
In \Cref{fig result}(a-b), we analyze a three-qubit model denoted as $|Q_sQ_hQ_l\rangle$. Here, $Q_s$ represents the spectator qubit, while $Q_h$ and $Q_l$ correspond to the high-frequency and low-frequency qubits involved in gates, respectively.
When the CZ gate involving $Q_h$ and $Q_l$ is not executed, the ZZ-coupling reaches its minimum value around 6.4 GHz. 
During the execution of the CZ gate, we have $\omega_{l}=\omega_{h}+\eta_{h}=\omega_\text{int}$. 
If $\omega_{c_{hs}}=6.4$ GHz, there will be a frequency range of about 100 MHz for $\omega_s$ where the CZ error is larger than $10^{-2}$. 
$Q_l$ and $Q_s$ are both coupled with $Q_h$. Considering the frequency crowding problem, $\omega_{s}$ is allocated around $\omega_\text{int}$ and likely to fall within the error large range. 
CAMEL uses compensation pulses to tune the $c_{hs}$ frequency from 6.4 GHz to 6.1 GHz, ensuring low error.

\begin{figure*}[htbp]
\centering
\begin{minipage}[t]{0.3\linewidth}
\centering
\includegraphics[width=\textwidth]{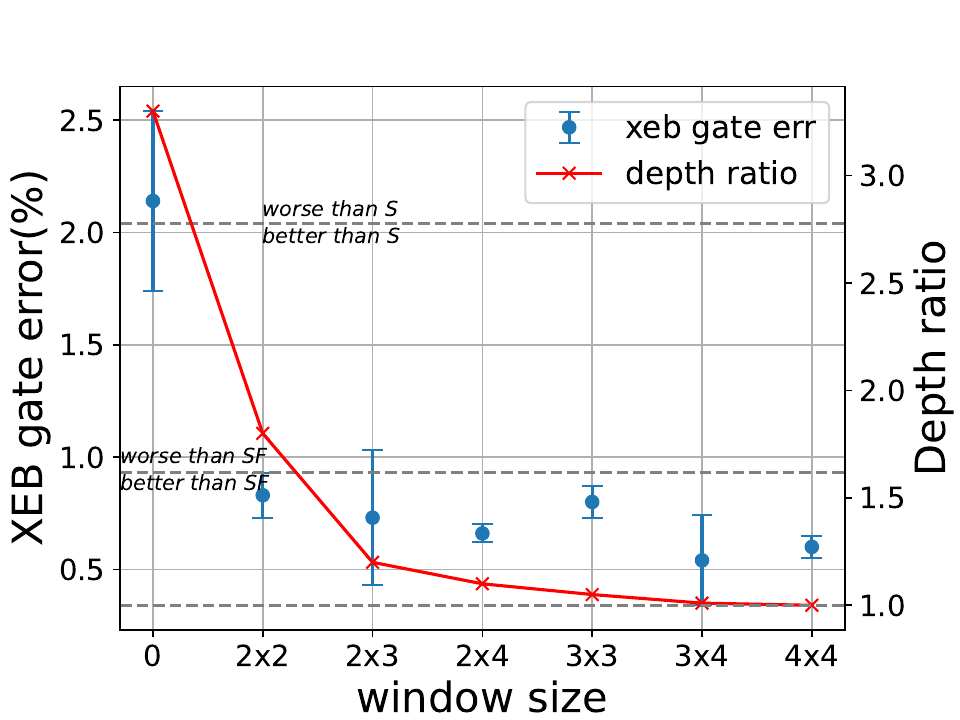}\\
(a)
\end{minipage}
\begin{minipage}[t]{0.3\linewidth}
\centering
\includegraphics[width=\textwidth]{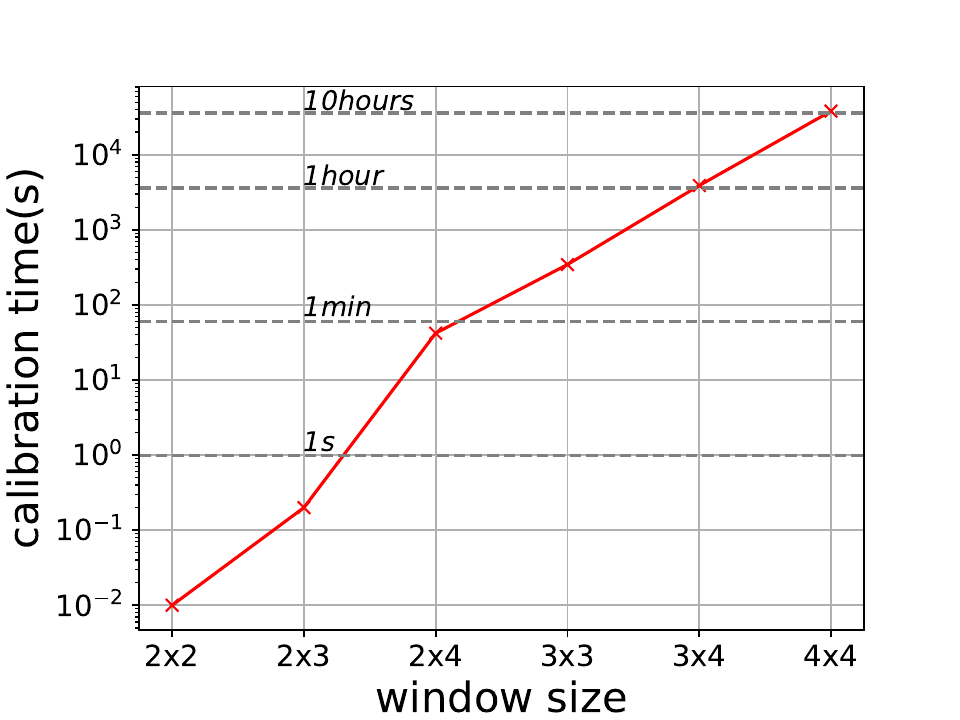}\\
(b)
\end{minipage}
\begin{minipage}[t]{0.3\linewidth}
\centering
\includegraphics[width=\textwidth]{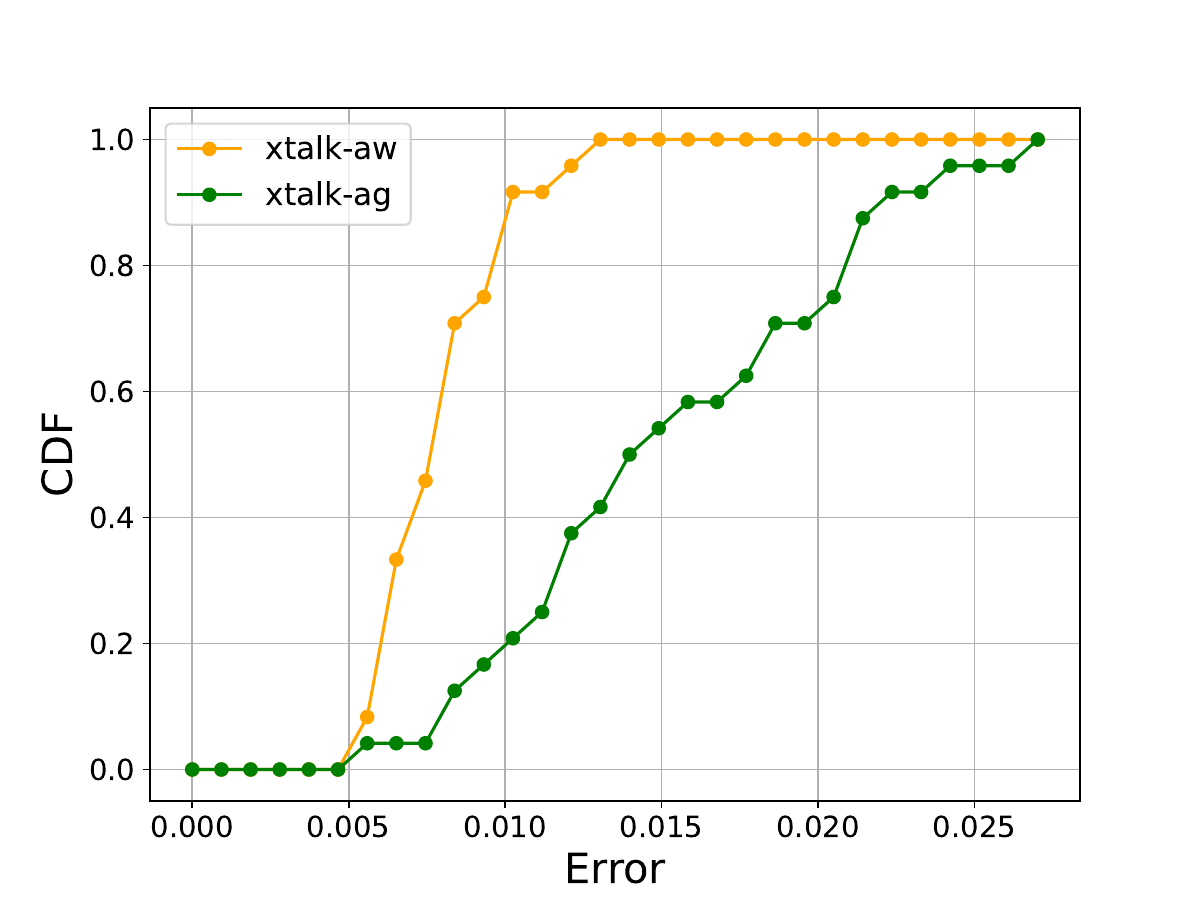}\\
(c)
\end{minipage}

\caption{
\REVISE{(a) The XEB error decreases as the window size increases from 0 to $4\times 4$. 
Starting at $2\times 2$, the error approaches the level of static frequency-aware compilation and continues to drop.
Additionally, the ratio of the circuit execution time after and before compilation decreases as the window size increases.
(b) The calibration time grows exponentially as window size increases.
(c) 
The cumulative distribution function (CDF) plot of errors from the XEB experiment.
The green curve represents the results compiled using the crosstalk-agnostic approach, 
while the yellow curve shows the results compiled using the crosstalk-aware mapper and scheduler. }
}
\label{fig windoweffwindow}
\end{figure*}

\subsubsection{Comparison with dynamic frequency-aware compilation}
CAMEL is better than the dynamic frequency-aware compilation baseline in terms of fidelity. 
Dynamic frequency-aware compilation requires dynamically allocating the interaction frequency for CZ gates activated by the algorithm subgraph. 
Otherwise, real-time frequency configuration, without gate parameter calibration, tends to result in low fidelity of CZ gates.
From \Cref{fig result}(c), when the low-frequency qubit is adjusted to $\omega_{l,\text{on}}$, the corresponding high-frequency qubit should be adjusted to $\omega_{l,\text{on}}-\eta_h$. 
However, at this frequency, there is no coupler frequency that can achieve an error lower than $10^{-1}$. 
Actually, there is a deviation between the set interaction frequency and the actual interaction frequency, necessitating calibration.
Implementing dynamic frequency configuration would require calibration before each activated algorithm subgraph, 
which would require an infeasible task of multiple iterations of gate parameter optimization during circuit execution.

\subsubsection{Comparison with serialization and static frequency-aware compilation}
CAMEL consistently outperforms the serialization compilation baseline in terms of fidelity. 
This can be understood by looking at \Cref{fig compilation}(a), which displays the depth ratio of compiled circuits.
The orange bars are higher than the gray dashed line, indicating that the compiled circuits take longer time to execute than CAMEL.
This occurs because the serialization baseline serializes all crosstalk gates, thereby increasing the execution time. Consequently, this amplifies the impact of decoherence.

Now we explain why static frequency-aware compilation baselines (green bars) have longer execution time.
Not all two-qubit gates can be executed simultaneously, and \Cref{fig compilation}(b-c) illustrates the eight maximum parallel patterns for the CZ gate. 
If a static frequency-aware compilation method is applied, with frequency configuration and calibration based on the first (last) four patterns, 
multiple CZ gates within any one of the ABCD (EFGH) patterns can be executed in parallel.
While a quantum circuit may necessitate various scenarios of parallelism, especially, parallel execution between CZ gates in distinct patterns, fixed-frequency configuration fails to accommodate such requirements.
Thus, the serialization method is still required in the presence of crosstalk, leading to an increase in execution time. 

\REVISE{
\subsubsection{Ablation study}
This section evaluates the effects of compensation pulses and the crosstalk-aware mapper and gate scheduler separately.

\textbf{With and without compensation pulse:}
Experiments with window sizes from 0 to $4\times 4$ are done in this step.
The baselines are serialization compilation the static frequency-aware compilation approaches. 
We use 200-layer, $4\times 4$ XEB circuits as benchmarks.
}

\REVISE{
When the window size is set to 0, CAMEL is equivalent to serialization compilation. 
As shown in \Cref{fig windoweffwindow}(a), CAMEL performs similarly to serialization compilation with a window size of 0.
However, as the window size increases, CAMEL demonstrates improved performance. Specifically, the XEB error decreases and the execution time ratio approaches 1, although the gains become marginal.
\Cref{fig windoweffwindow}(b) illustrates that the calibration time grows exponentially with the window size. 
At a window size of $2\times 2$, the XEB error is already lower than that of static frequency-aware compilation.
This supports the choice of a $2\times 2$ window as an optimal balance between fidelity and calibration time.

Within the calibrated window, all parallel two-qubit gate situations are considered, and all frequency crowding scenarios can be mitigated by compensation pulses. 
This window size introduces maximum allowable parallelism in the subsequent mapping and scheduling steps, increasing the solution space for the mapping and scheduling approach, 
making it easier to find an optimal solution.
}

\REVISE{
\textbf{With and without mapping and scheduling:}
In this step, we conduct a control experiment comparing the crosstalk-agnostic approach with the crosstalk-aware mapper and scheduler (the full CAMEL) on a chip after frequency configuration and pulse calibration.
We performed frequency configuration based on the ABCD coupler activation pattern and calibrated the compensation pulse for each $2\times 2$ window. 
An XEB experiment with four randomly selected maximum parallel coupler activation patterns were conducted.
}

% \begin{figure}[htbp]
% \centering
% \includegraphics[width=0.4\textwidth]{compileeff1.pdf}\\
% \caption{
% \REVISE{
% The green curve represents the results compiled using the crosstalk-agnostic approach, 
% while the yellow curve shows the results compiled using the crosstalk-aware mapper and scheduler. 
% The cumulative distribution function (CDF) plot of errors from the XEB experiment using random coupler activation patterns.}
% }
% \label{fig conpileeff}
% \end{figure}

\REVISE{As shown in \Cref{fig windoweffwindow}(c)), the error distribution of CAMEL is lower than that of the crosstalk-agnostic approach. This is because the crosstalk-agnostic approach fails to account for frequency crowding and, when encountering cases where compensation pulses cannot fully address the crowding, it does not optimize the mapping or scheduling. 
Without mapping and scheduling, for the window size is much smaller than the chip size, frequency crowding can only be mitigated locally. 
In contrast, CAMEL’s mapper and scheduler are capable of detecting and mitigating frequency crowding in arbitrary quantum circuits.
The CAMEL mapping and scheduling components use a heuristic approach to extend the local crosstalk mitigation capability of the compensation pulse to the entire chip.}

\section{Conclusion}
% In summary, we have proposed a compilation approach for crosstalk and decoherence mitigation in superconducting frequency-tunable quantum chips. 
% We first validated numerically that applying compensation pulse to the spectator coupler effectively mitigates crosstalk, especially in situation where frequency crowding occurs. 
% To address the challenge of optimizing compensating pulse for arbitrary parallel patterns, we introduced a sliding window approach. 
% Based on the compensation pulse, we devised an optimization mapping considering crosstalk and a gate timing scheduling approach. 
% Through numerical experiments and comparisons with existing frequency configuration compilation approaches, CAMEL demonstrates reduced crosstalk and shorter execution times for common NISQ benchmark circuits on lattice-structured quantum chips.
% CAMEL holds significant promise for the development of robust and scalable quantum computing systems and lays the groundwork for future large-scale quantum error correction circuits that require parallel execution of multiple CZ gates.
In summary, we propose a compilation approach to mitigate crosstalk and decoherence in superconducting frequency-tunable quantum chips. Our method first numerically validates that applying compensation pulses to the spectator coupler effectively reduces crosstalk, particularly in scenarios where frequency crowding occurs. To tackle the challenge of optimizing compensation pulses for arbitrary parallel patterns, we introduced a sliding window approach. Building on this, we developed a crosstalk-aware qubit mapping strategy and a gate timing scheduling method named ``CAMEL''. Through numerical experiments and comparisons with existing frequency configuration compilation methods, CAMEL shows notable improvements in reducing crosstalk and achieving shorter execution times for common NISQ benchmark circuits on lattice-structured superconducting quantum chips. CAMEL presents a promising step toward developing robust and scalable quantum computing systems, providing a foundation for future large-scale quantum error correction circuits that require the parallel execution of multiple CZ gates.

\section*{Acknowledgements}
This work has been supported by the National Key Research and Development Program of China (Grant No. 2023YFB4502500).
%%%%%%% -- PAPER CONTENT ENDS -- %%%%%%%%

%%%%%%%%% -- BIB STYLE AND FILE -- %%%%%%%%
% \bibliographystyle{IEEEtranS}
\bibliographystyle{unsrt}
\bibliography{refs}

\begin{thebibliography}{10}

\bibitem{preskill2018quantum}
John Preskill.
\newblock Quantum computing in the nisq era and beyond.
\newblock {\em Quantum}, 2:79, 2018.

\bibitem{krantz2019quantum}
Philip Krantz, Morten Kjaergaard, Fei Yan, Terry~P Orlando, Simon Gustavsson, and William~D Oliver.
\newblock A quantum engineer's guide to superconducting qubits.
\newblock {\em Applied physics reviews}, 6(2), 2019.

\bibitem{bylander2011noise}
Jonas Bylander, Simon Gustavsson, Fei Yan, Fumiki Yoshihara, Khalil Harrabi, George Fitch, David~G Cory, Yasunobu Nakamura, Jaw-Shen Tsai, and William~D Oliver.
\newblock Noise spectroscopy through dynamical decoupling with a superconducting flux qubit.
\newblock {\em Nature Physics}, 7(7):565--570, 2011.

\bibitem{ithier2005decoherence}
G~Ithier, E~Collin, P~Joyez, PJ~Meeson, Denis Vion, Daniel Esteve, F~Chiarello, A~Shnirman, Yu~Makhlin, Josef Schriefl, et~al.
\newblock Decoherence in a superconducting quantum bit circuit.
\newblock {\em Physical Review B}, 72(13):134519, 2005.

\bibitem{yan2016flux}
Fei Yan, Simon Gustavsson, Archana Kamal, Jeffrey Birenbaum, Adam~P Sears, David Hover, Ted~J Gudmundsen, Danna Rosenberg, Gabriel Samach, Steven Weber, et~al.
\newblock The flux qubit revisited to enhance coherence and reproducibility.
\newblock {\em Nature communications}, 7(1):12964, 2016.

\bibitem{krinner2020benchmarking}
Sebastian Krinner, Stefania Lazar, Ants Remm, Christian~K Andersen, Nathan Lacroix, Graham~J Norris, Christoph Hellings, Mihai Gabureac, Christopher Eichler, and Andreas Wallraff.
\newblock Benchmarking coherent errors in controlled-phase gates due to spectator qubits.
\newblock {\em Physical Review Applied}, 14(2):024042, 2020.

\bibitem{almudever2017engineering}
Carmen~G Almudever, Lingling Lao, Xiang Fu, Nader Khammassi, Imran Ashraf, Dan Iorga, Savvas Varsamopoulos, Christopher Eichler, Andreas Wallraff, Lotte Geck, et~al.
\newblock The engineering challenges in quantum computing.
\newblock In {\em Design, Automation \& Test in Europe Conference \& Exhibition (DATE), 2017}, pages 836--845. IEEE, 2017.

\bibitem{siraichi2019qubit}
Marcos~Yukio Siraichi, Vin{\'\i}cius Fernandes~dos Santos, Caroline Collange, and Fernando Magno~Quint{\~a}o Pereira.
\newblock Qubit allocation as a combination of subgraph isomorphism and token swapping.
\newblock {\em Proceedings of the ACM on Programming Languages}, 3(OOPSLA):1--29, 2019.

\bibitem{siraichi2018qubit}
Marcos~Yukio Siraichi, Vin{\'\i}cius Fernandes~dos Santos, Caroline Collange, and Fernando Magno~Quint{\~a}o Pereira.
\newblock Qubit allocation.
\newblock In {\em Proceedings of the 2018 International Symposium on Code Generation and Optimization}, pages 113--125, 2018.

\bibitem{murali2020software}
Prakash Murali, David~C McKay, Margaret Martonosi, and Ali Javadi-Abhari.
\newblock Software mitigation of crosstalk on noisy intermediate-scale quantum computers.
\newblock In {\em Proceedings of the Twenty-Fifth International Conference on Architectural Support for Programming Languages and Operating Systems}, pages 1001--1016, 2020.

\bibitem{Hua2022CQCAC}
Fei Hua, Yuwei Jin, Yan-Hao Chen, Chi Zhang, Ari~B. Hayes, Hang Gao, and Eddy~Z. Zhang.
\newblock Cqc: A crosstalk-aware quantum program compilation framework.
\newblock 2022.

\bibitem{finck2021suppressed}
ADK Finck, S~Carnevale, D~Klaus, Chris Scerbo, J~Blair, TG~McConkey, Cihan Kurter, A~Carniol, George Keefe, Muir Kumph, et~al.
\newblock Suppressed crosstalk between two-junction superconducting qubits with mode-selective exchange coupling.
\newblock {\em Physical Review Applied}, 16(5):054041, 2021.

\bibitem{ku2020suppression}
Jaseung Ku, Xuexin Xu, Markus Brink, David~C McKay, Jared~B Hertzberg, Mohammad~H Ansari, and BLT Plourde.
\newblock Suppression of unwanted z z interactions in a hybrid two-qubit system.
\newblock {\em Physical review letters}, 125(20):200504, 2020.

\bibitem{ni2022scalable}
Zhongchu Ni, Sai Li, Libo Zhang, Ji~Chu, Jingjing Niu, Tongxing Yan, Xiuhao Deng, Ling Hu, Jian Li, Youpeng Zhong, et~al.
\newblock Scalable method for eliminating residual z z interaction between superconducting qubits.
\newblock {\em Physical review letters}, 129(4):040502, 2022.

\bibitem{wei2021quantum}
KX~Wei, E~Magesan, I~Lauer, S~Srinivasan, DF~Bogorin, S~Carnevale, GA~Keefe, Y~Kim, D~Klaus, W~Landers, et~al.
\newblock Quantum crosstalk cancellation for fast entangling gates and improved multi-qubit performance.
\newblock {\em arXiv preprint arXiv:2106.00675}, 2021.

\bibitem{zhao2021suppression}
Peng Zhao, Dong Lan, Peng Xu, Guangming Xue, Mace Blank, Xinsheng Tan, Haifeng Yu, and Yang Yu.
\newblock Suppression of static z z interaction in an all-transmon quantum processor.
\newblock {\em Physical Review Applied}, 16(2):024037, 2021.

\bibitem{mundada2019suppression}
Pranav Mundada, Gengyan Zhang, Thomas Hazard, and Andrew Houck.
\newblock Suppression of qubit crosstalk in a tunable coupling superconducting circuit.
\newblock {\em Physical Review Applied}, 12(5):054023, 2019.

\bibitem{sung2021realization}
Youngkyu Sung, Leon Ding, Jochen Braum{\"u}ller, Antti Veps{\"a}l{\"a}inen, Bharath Kannan, Morten Kjaergaard, Ami Greene, Gabriel~O Samach, Chris McNally, David Kim, et~al.
\newblock Realization of high-fidelity cz and z z-free iswap gates with a tunable coupler.
\newblock {\em Physical Review X}, 11(2):021058, 2021.

\bibitem{klimov2024optimizing}
Paul~V Klimov, Andreas Bengtsson, Chris Quintana, Alexandre Bourassa, Sabrina Hong, Andrew Dunsworth, Kevin~J Satzinger, William~P Livingston, Volodymyr Sivak, Murphy~Yuezhen Niu, et~al.
\newblock Optimizing quantum gates towards the scale of logical qubits.
\newblock {\em Nature Communications}, 15(1):2442, 2024.

\bibitem{Ding2020SystematicCM}
Yongshan Ding, Pranav Gokhale, Sophia~Fuhui Lin, Rich Rines, Thomas~P. Propson, and Frederic~T. Chong.
\newblock Systematic crosstalk mitigation for superconducting qubits via frequency-aware compilation.
\newblock {\em 2020 53rd Annual IEEE/ACM International Symposium on Microarchitecture (MICRO)}, pages 201--214, 2020.

\bibitem{wittler2021integrated}
Nicolas Wittler, Federico Roy, Kevin Pack, Max Werninghaus, Anurag~Saha Roy, Daniel~J Egger, Stefan Filipp, Frank~K Wilhelm, and Shai Machnes.
\newblock Integrated tool set for control, calibration, and characterization of quantum devices applied to superconducting qubits.
\newblock {\em Physical review applied}, 15(3):034080, 2021.

\bibitem{klimov2020snake}
Paul~V Klimov, Julian Kelly, John~M Martinis, and Hartmut Neven.
\newblock The snake optimizer for learning quantum processor control parameters.
\newblock {\em arXiv preprint arXiv:2006.04594}, 2020.

\bibitem{arute2019quantum}
Frank Arute, Kunal Arya, Ryan Babbush, Dave Bacon, Joseph~C Bardin, Rami Barends, Rupak Biswas, Sergio Boixo, Fernando~GSL Brandao, David~A Buell, et~al.
\newblock Quantum supremacy using a programmable superconducting processor.
\newblock {\em Nature}, 574(7779):505--510, 2019.

\bibitem{Clarktdag}
Joseph Clark, Travis Humble, and Himanshu Thapliyal.
\newblock Tdag: Tree-based directed acyclic graph partitioning for quantum circuits.
\newblock In {\em Proceedings of the Great Lakes Symposium on VLSI 2023}, GLSVLSI '23, page 587–592, New York, NY, USA, 2023. Association for Computing Machinery.

\bibitem{muller2015interacting}
Clemens M{\"u}ller, J{\"u}rgen Lisenfeld, Alexander Shnirman, and Stefano Poletto.
\newblock Interacting two-level defects as sources of fluctuating high-frequency noise in superconducting circuits.
\newblock {\em Physical Review B}, 92(3):035442, 2015.

\bibitem{bravyi2011schrieffer}
Sergey Bravyi, David~P DiVincenzo, and Daniel Loss.
\newblock Schrieffer--wolff transformation for quantum many-body systems.
\newblock {\em Annals of physics}, 326(10):2793--2826, 2011.

\bibitem{2011Implementing}
Matteo Mariantoni, H.~Wang, T.~Yamamoto, M.~Neeley, and John~M Martinis.
\newblock Implementing the quantum von neumann architecture with superconducting circuits.
\newblock {\em Science}, 334(6052):61--65, 2011.

\bibitem{2023Generation}
Sirui Cao, Bujiao Wu, Fusheng Chen, Ming Gong, Yulin Wu, Yangsen Ye, Chen Zha, Haoran Qian, Chong Ying, and Shaojun Guo.
\newblock Generation of genuine entanglement up to 51 superconducting qubits.
\newblock {\em Nature}, 2023.

\bibitem{google2023suppressing}
Suppressing quantum errors by scaling a surface code logical qubit.
\newblock {\em Nature}, 614(7949):676--681, 2023.

\bibitem{acharya2024quantum}
Rajeev Acharya, Laleh Aghababaie-Beni, Igor Aleiner, Trond~I Andersen, Markus Ansmann, Frank Arute, Kunal Arya, Abraham Asfaw, Nikita Astrakhantsev, Juan Atalaya, et~al.
\newblock Quantum error correction below the surface code threshold.
\newblock {\em arXiv preprint arXiv:2408.13687}, 2024.

\bibitem{zhao2022quantum}
Peng Zhao, Kehuan Linghu, Zhiyuan Li, Peng Xu, Ruixia Wang, Guangming Xue, Yirong Jin, and Haifeng Yu.
\newblock Quantum crosstalk analysis for simultaneous gate operations on superconducting qubits.
\newblock {\em PRX quantum}, 3(2):020301, 2022.

\bibitem{zhao2022spurious}
Peng Zhao, Yingshan Zhang, Xuegang Li, Jiaxiu Han, Huikai Xu, Guangming Xue, Yirong Jin, and Haifeng Yu.
\newblock Spurious microwave crosstalk in floating superconducting circuits.
\newblock {\em arXiv preprint arXiv:2206.03710}, 2022.

\bibitem{Xie2022SuppressingZC}
Lei Xie, Jidong Zhai, Zhenxing Zhang, Jonathan Allcock, Shengyu Zhang, and Yicong Zheng.
\newblock Suppressing zz crosstalk of quantum computers through pulse and scheduling co-optimization.
\newblock {\em Proceedings of the 27th ACM International Conference on Architectural Support for Programming Languages and Operating Systems}, 2022.

\bibitem{klimov2020optimizing}
Paul Klimov, Julian Kelly, Kevin Satzinger, Zijun Chen, Hartmut Neven, and John Martinis.
\newblock Optimizing quantum gate frequencies for google’s quantum processors.
\newblock {\em Bulletin of the American Physical Society}, 65, 2020.

\bibitem{chu2021coupler}
Ji~Chu and Fei Yan.
\newblock Coupler-assisted controlled-phase gate with enhanced adiabaticity.
\newblock {\em Physical Review Applied}, 16(5):054020, 2021.

\bibitem{zajac2021spectator}
DM~Zajac, J~Stehlik, DL~Underwood, T~Phung, J~Blair, S~Carnevale, D~Klaus, GA~Keefe, A~Carniol, M~Kumph, et~al.
\newblock Spectator errors in tunable coupling architectures.
\newblock {\em arXiv preprint arXiv:2108.11221}, 2021.

\bibitem{samotij2015counting}
Wojciech Samotij.
\newblock Counting independent sets in graphs.
\newblock {\em European journal of combinatorics}, 48:5--18, 2015.

\bibitem{jou2000number}
Min-Jen Jou and Gerard~J Chang.
\newblock The number of maximum independent sets in graphs.
\newblock {\em Taiwanese Journal of Mathematics}, 4(4):685--695, 2000.

\bibitem{shindi2023model}
Omar Shindi, Qi~Yu, Parth Girdhar, and Daoyi Dong.
\newblock Model-free quantum gate design and calibration using deep reinforcement learning.
\newblock {\em IEEE Transactions on Artificial Intelligence}, 2023.

\bibitem{hagberg2020networkx}
Aric Hagberg and Drew Conway.
\newblock Networkx: Network analysis with python.
\newblock {\em URL: https://networkx. github. io}, 2020.

\bibitem{adedoyin2018quantum}
Adetokunbo Adedoyin, John Ambrosiano, Petr Anisimov, William Casper, Gopinath Chennupati, Carleton Coffrin, Hristo Djidjev, David Gunter, Satish Karra, Nathan Lemons, et~al.
\newblock Quantum algorithm implementations for beginners.
\newblock {\em arXiv preprint arXiv:1804.03719}, 2018.

\bibitem{zhang2021time}
Chi Zhang, Ari~B Hayes, Longfei Qiu, Yuwei Jin, Yanhao Chen, and Eddy~Z Zhang.
\newblock Time-optimal qubit mapping.
\newblock In {\em Proceedings of the 26th ACM International Conference on Architectural Support for Programming Languages and Operating Systems}, pages 360--374, 2021.

\bibitem{venturelli2017temporal}
Davide Venturelli, Minh Do, Eleanor Rieffel, and Jeremy Frank.
\newblock Temporal planning for compilation of quantum approximate optimization circuits.
\newblock In {\em Scheduling and Planning Applications woRKshop (SPARK)}, page~58, 2017.

\bibitem{zulehner2018efficient}
Alwin Zulehner, Alexandru Paler, and Robert Wille.
\newblock An efficient methodology for mapping quantum circuits to the ibm qx architectures.
\newblock {\em IEEE Transactions on Computer-Aided Design of Integrated Circuits and Systems}, 38(7):1226--1236, 2018.

\bibitem{li2019tackling}
Gushu Li, Yufei Ding, and Yuan Xie.
\newblock Tackling the qubit mapping problem for nisq-era quantum devices.
\newblock In {\em Proceedings of the Twenty-Fourth International Conference on Architectural Support for Programming Languages and Operating Systems}, pages 1001--1014, 2019.

\bibitem{kjaergaard2020programming}
Morten Kjaergaard, Mollie~E Schwartz, Ami Greene, Gabriel~O Samach, Andreas Bengtsson, Michael O'Keeffe, Christopher~M McNally, Jochen Braum{\"u}ller, David~K Kim, Philip Krantz, et~al.
\newblock Programming a quantum computer with quantum instructions.
\newblock {\em arXiv preprint arXiv:2001.08838}, 2020.

\bibitem{choi2019tutorial}
Jaeho Choi and Joongheon Kim.
\newblock A tutorial on quantum approximate optimization algorithm (qaoa): Fundamentals and applications.
\newblock In {\em 2019 International Conference on Information and Communication Technology Convergence (ICTC)}, pages 138--142. IEEE, 2019.

\bibitem{johansson2012qutip}
J~Robert Johansson, Paul~D Nation, and Franco Nori.
\newblock Qutip: An open-source python framework for the dynamics of open quantum systems.
\newblock {\em Computer Physics Communications}, 183(8):1760--1772, 2012.

\end{thebibliography}
%%%%%%%%%%%%%%%%%%%%%%%%%%%%%%%%%%%%

\begin{IEEEbiography}[{\includegraphics[width=1in,height=1.25in,clip,keepaspectratio]{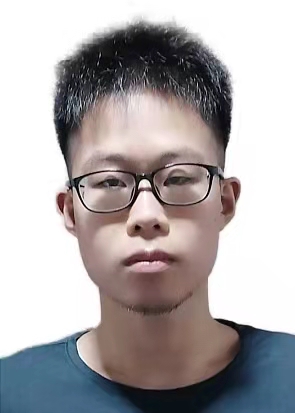}}]{Bin-Han Lu}
Bin-Han Lu was born in Guangzhou, China, in 1997. He earned his Bachelor’s degree from Jinan University in 2019. Following that, he has been pursuing a Ph.D. at the CAS Center for Excellence in Quantum Information and Quantum Physics, University of Science and Technology of China, from 2019 to 2024. His research interests include quantum computing, quantum algorithms, and quantum hardware.
\end{IEEEbiography}

\begin{IEEEbiography}[{\includegraphics[width=1in,height=1.25in,clip,keepaspectratio]{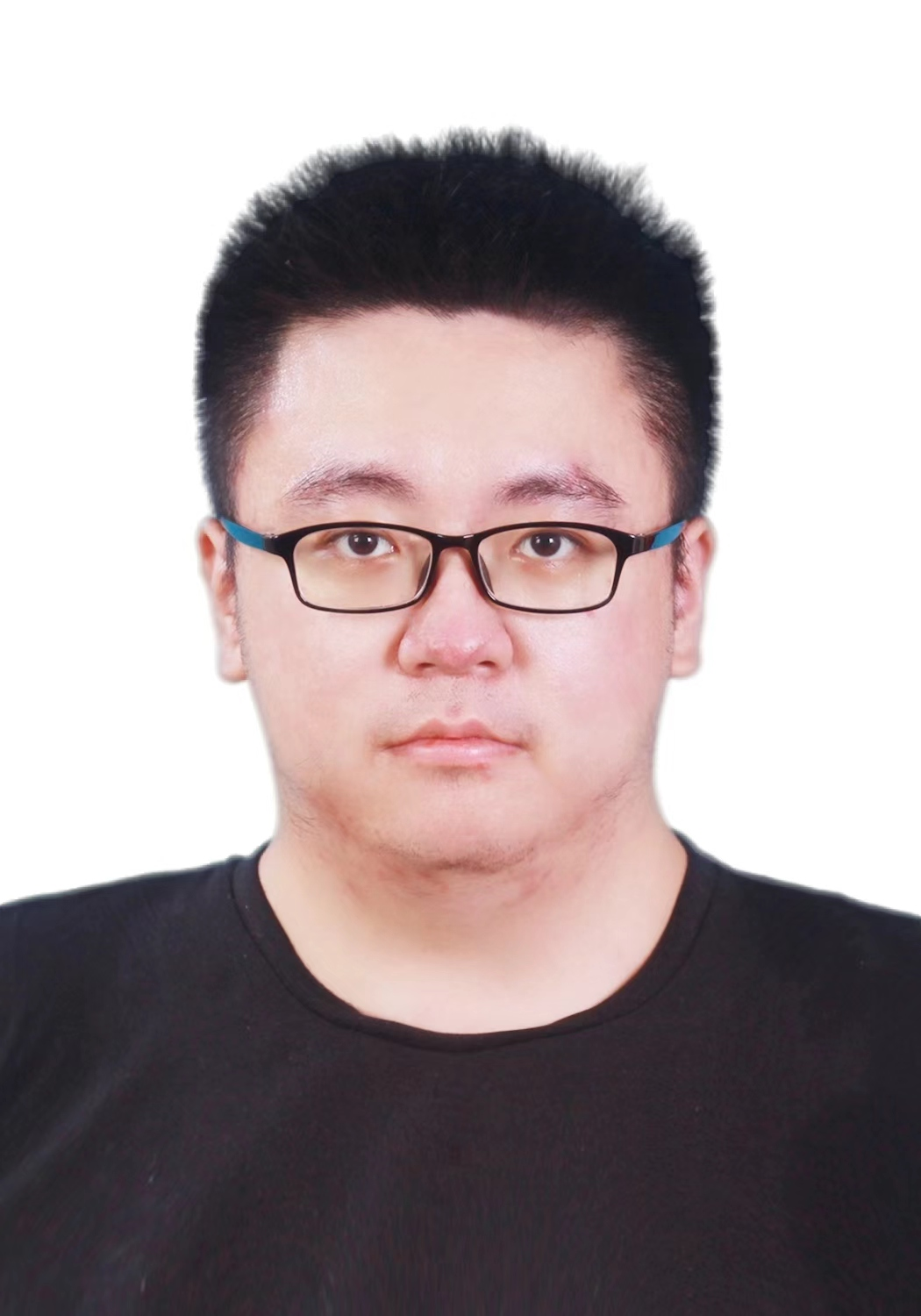}}]{Zhao-Yun Chen}
Zhao-Yun Chen was born in Wuhan, China, in 1994. He got his Bachelor’s degree from the University of Science and Technology of China (USTC) in 2016 and his Ph.D. in Physics from USTC in 2021. After that, he worked as a postdoctoral researcher at the Institute of Artificial Intelligence, Hefei Comprehensive National Science Center from 2021 to 2024. He then became an associate researcher there. His research interests include quantum computing, quantum algorithms, and quantum software.
\end{IEEEbiography}

\begin{IEEEbiography}[{\includegraphics[width=1in,height=1.25in,clip,keepaspectratio]{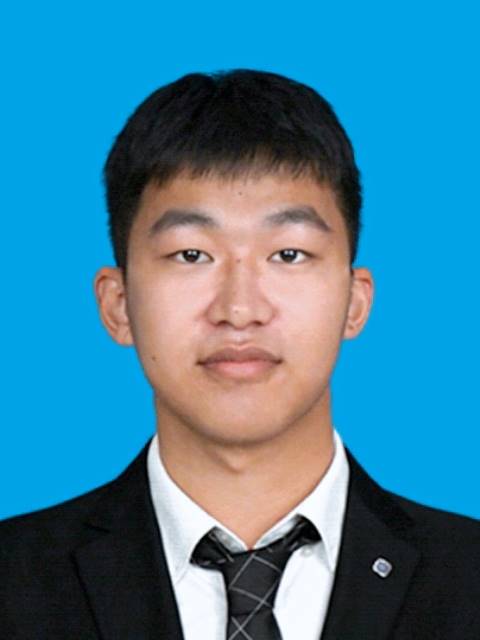}}]{Peng Wang}
Peng Wang received the B.S. degree in physics from Ocean University of China, Qingdao, China, in 2021. He is currently pursuing the Ph.D. degree at CAS Key Laboratory of Quantum Information, University of Science and Technology of China. His research interests include superconducting quantum computing, quantum algorithms, and quantum simulation.
\end{IEEEbiography}

\begin{IEEEbiography}[{\includegraphics[width=1in,height=1.25in,clip,keepaspectratio]{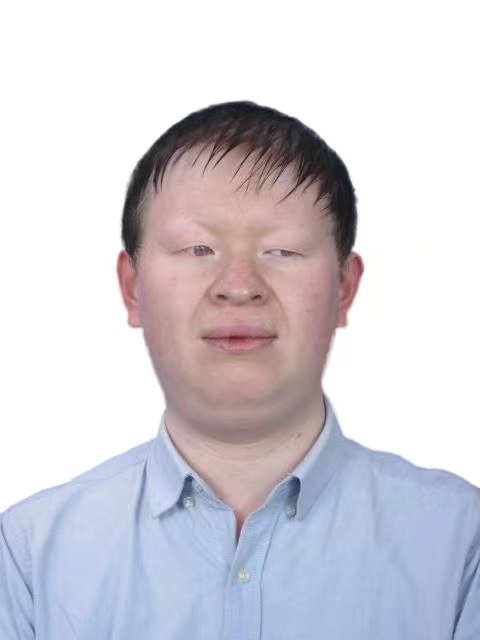}}]{Huan-Yu Liu}
Huan-Yu Liu is from Fuyang, Anhui, China. He got his Bachelor's degree from Hefei University of Technology in 2018 and his Ph.D. from the University of Science and Technology of China in 2023. He is currently engaged in postdoctoral research at the University of Science and Technology of China. His main research interests include quantum computing, quantum algorithms, and quantum simulation.
\end{IEEEbiography}

\begin{IEEEbiography}[{\includegraphics[width=1in,height=1.25in,clip,keepaspectratio]{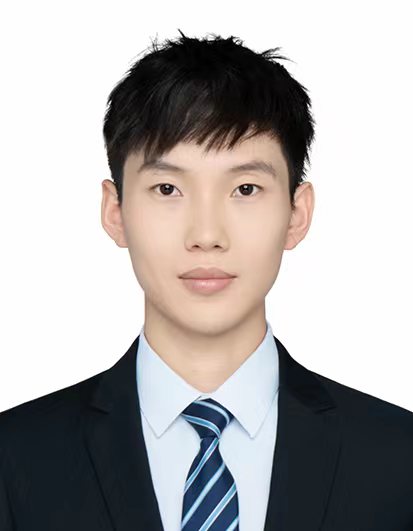}}]{Tai-Ping Sun}
Tai-Ping Sun received the B.S. degree in Material Physics from University of Science and Technology of China, Hefei, China, in 2019. He is currently working toward the Ph.D. degree in Physics with the Department of Physics, University of Science and Technology of China, Hefei, China. His research interests include quantum computation, quantum machine learning and time series processing.
\end{IEEEbiography}

\begin{IEEEbiography}[{\includegraphics[width=1in,height=1.25in,clip,keepaspectratio]{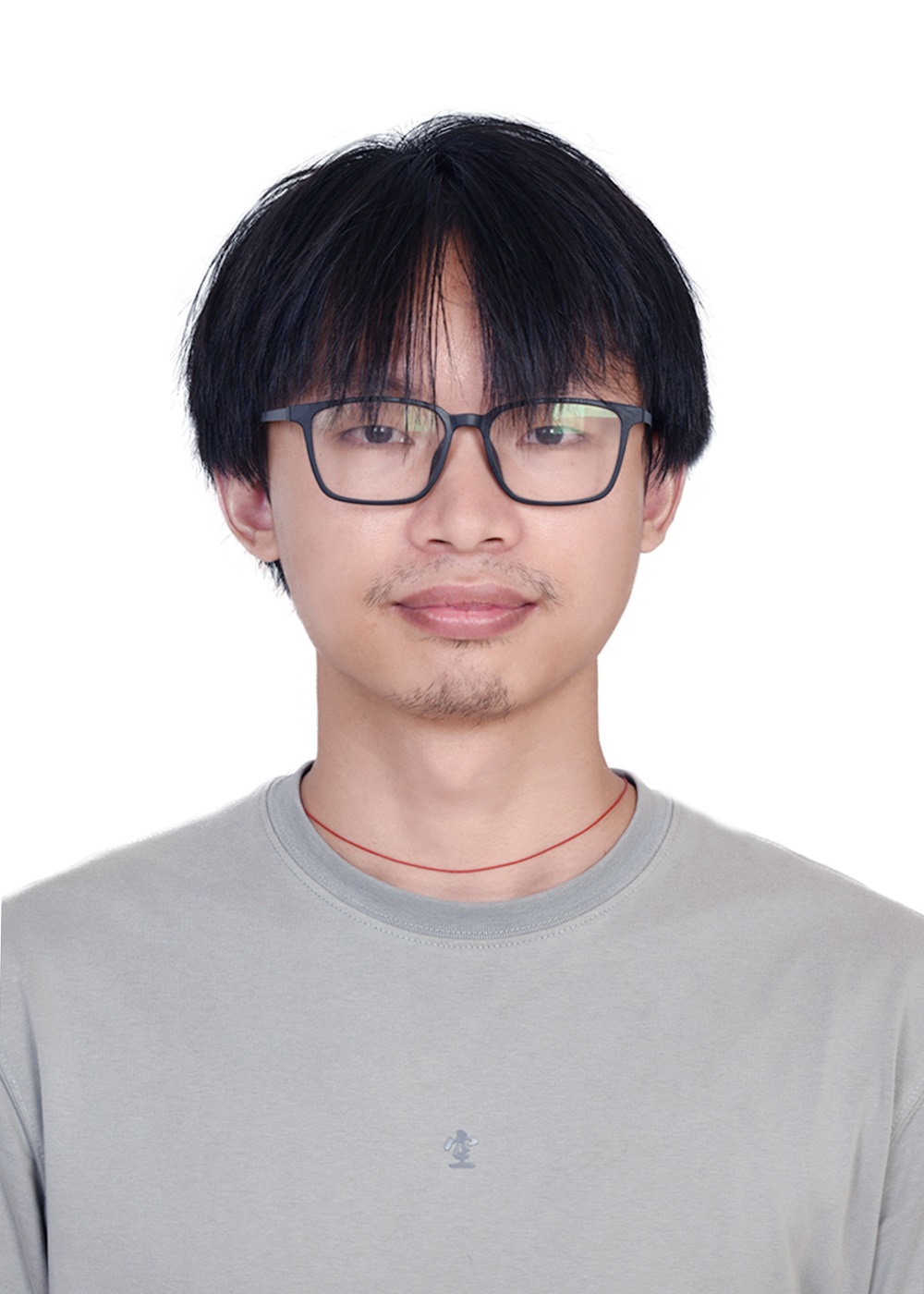}}]{Peng Duan}
Peng Duan was  born in Jiujiang, China, in 1993. He got his Bachelor's degree from Shandong University in 2015 and his Ph.D. in physics from University of Science and Technology of China (USTC) in 2022. He is now a postdoctoral researcher at USTC. His main research interests focus on quantum computing and superconducting quantum devices.
\end{IEEEbiography}

\begin{IEEEbiography}[{\includegraphics[width=1in,height=1.25in,clip,keepaspectratio]{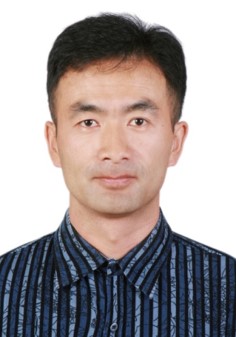}}]{Yu-Chun Wu}
Yu-Chun Wu, Ph.D., Assoc. Prof. Born in Qinghai, May 1974. Entered Shaanxi Normal University Math Dept. in 1991, graduated with a Master’s in 1998. Received a Ph.D. from the CAS Institute of Mathematics and Systems Science in 2001. Postdoc at USTC Quantum Information Lab from Jul 2001 to Feb 2004, stayed on as faculty. Additional postdoc at Gdansk University, Poland from Apr 2007 to Dec 2008. Research interests include quantum information theory, non-locality, entanglement, and quantum correlations.
\end{IEEEbiography}

\begin{IEEEbiography}[{\includegraphics[width=1in,height=1.25in,clip,keepaspectratio]{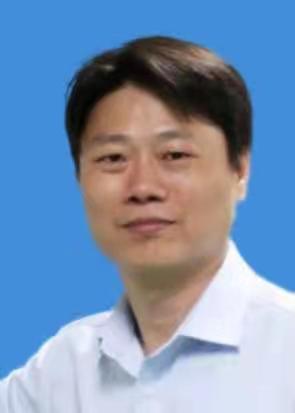}}]{Guo-Ping Guo}
Guo-Ping Guo, born in Nanchang, Jiangxi in December 1977, is a professor and doctoral supervisor. He is the leader in the research direction of semiconductor quantum dot quantum chips in the laboratory. He is the chief scientist of National Key Basic Research and Development Program Project Class A (Super 973) and a recipient of the National Distinguished Young Scholars Fund.
\end{IEEEbiography}
\end{document}